\RequirePackage{snapshot} 

\documentclass[12pt]{elsarticle}

\usepackage{graphicx}
\usepackage{epstopdf}
\usepackage{amsfonts}
\usepackage[english]{babel}
\usepackage{amsmath}
\usepackage{bm}
\RequirePackage[colorlinks,citecolor=blue,urlcolor=blue]{hyperref}
\usepackage{enumerate}
\usepackage{fullpage}
\usepackage{natbib}
\setcitestyle{authoryear,open={(},close={)}}
\DeclareRobustCommand{\VAN}[3]{#2}
\newtheorem{theorem}{Theorem}

\newtheorem{proposition}[theorem]{Proposition}
\newtheorem{remark}[theorem]{Remark}
\graphicspath{ {./immagini/} }

\begin{document}

\begin{frontmatter}
	
	
	
	\title{Classification methods for Hilbert data\\ based on surrogate density}
	
	\author{Enea~G.~Bongiorno\footnote{​Dipartimento di Studi per l'Economia e l'Impresa, Universit\`{a} del Piemonte Orientale, Via Perrone 18,	28100, Novara, Italia. Tel: (+39) 0321 375 317 e--mail: enea.bongiorno@uniupo.it}, Aldo~Goia}
	\address{Universit\`{a} del Piemonte Orientale, Italy}
	
		
	\begin{abstract}
	An unsupervised and a supervised classification approaches for Hilbert random curves are studied. Both rest on the use of a surrogate of the probability density which is defined, in a distribution-free mixture context, from an asymptotic factorization of the small-ball probability. That surrogate density is estimated by a kernel approach from the principal components of the data. The focus is on the illustration of the classification algorithms and the computational implications, with particular attention to the tuning of the parameters involved. Some asymptotic results are sketched. Applications on simulated and real datasets show how the proposed methods work.
	\end{abstract}
	
	
\end{frontmatter}

\section*{Introduction}

In multivariate classification problems, whether they are supervised or unsupervised, joint density, or better its estimate, plays a central role. In order to make it clear, one has just to think about the literature on the model based clustering approaches, and recall that all the discriminant methods resting on the so-called Bayes rule require an estimation of the joint density in each group (for recent developments, see for instance \citealp{boc:ing:ver13,boc:ing:ver14,gim:han:imi12}).

When one deals with data belonging to functional spaces (for a general introduction on this topic, one can refer to the monographs of \citealp{fer:vie06}, \citealp{hor:kok12} and \citealp{ram:sil05}, and to the recent book \citealp{bon:goi:sal:vie14}), the dimensionality problem arises immediately, and, as a consequence, a probability density function generally does not exist (see \citealp{del:hal10}). Hence, a direct extension of density oriented classical multivariate classification approaches to functional data cannot be implemented: usually, a reduction of dimensionality based on projection over suitable finite subspaces, is made as a preliminary step to tackle the problem. This route is followed for instance by \cite{jam:sug03}, where model based clustering methods are proposed, and by \cite{jam:has01} and \cite{shi08} in defining a discriminant approach: the general idea is to put a suitable density mixture model over the coefficients of the representation of functional data in the finite subspace, admitting that such model may summarize the distribution of the underlying process. It is worth noting that also other dimensionality reduction approaches are possible: for instance, one can recall the techniques, which rest on the most important points, illustrated in \cite{del:hal:bat12}.

Another way to proceed, that aims at working directly on the distribution of the process, refers to the concept of surrogate (or pseudo) density. The general principle, dating back to \cite{gas:hal:pre98}, is to factorize the small ball probability associated with the functional data, when the radius of the ball tends to zero, as a product of two terms: an ``intensity term'' which depends only on the center of the ball, and a kind of ``volume parameter'' which depends only on the radius. Since the first term reflects the latent structure of the distribution of the underlying process, it represents an ideal candidate to play the role that the multivariate density has in finite dimensional classification methods. Theoretical conditions that allow such a factorization when one considers Hilbert functional data in the space determined by the basis of the Karhunen--Lo\`{e}ve decomposition (i.e.~by the eigenfunctions of the principal components analysis), and the structure of the pseudo-density (linked to the principal components, i.e.~the coefficients of the decomposition), are discussed in \cite{del:hal10} and then in \cite{bon:goi15}, where some assumptions are relaxed. One can observe that this approach allows to see the projective approaches in a more general theoretical context.

To put into effect the above factorization and take advantage of the pseudo-density, different ways are possible. A first one is to specify a suitable density model mixture for the principal components: this full parametric approach is followed by \cite{jac:pre14} in defining a Gaussian mixture clustering procedure. On the other hand, a full nonparametric approach is possible as done in \cite{fer:kud:vie12} where a k--NN procedure is proposed to estimate the pseudo-density. 

In this paper we consider an intermediate approach for evaluating the
pseudo-density: after computing the first $d$ principal components of the
functional data, we obtain an estimate of their joint density $f_{d}$ via
the classical Parzen--Rosenblatt kernel method. We can see this approach as
semi-parametric: if on one hand, we use coefficients of the Karhunen--Lo\`{e}ve decomposition in defining the pseudo-density (it is not estimated
directly), on the other hand, the mixture model is not specified. 

The goal of this paper is to clarify how to use the proposed method to
tackle classification problems for Hilbert data: we illustrate both a pseudo-density oriented clustering algorithm resting on the definition of clusters as a high intensity regions from the modes of $f_{d}$, and a classifier based on the Bayes rule in the discriminant context, under suitable hypothesis on the distributions of the mixture components. After introducing the general mixture model and the theoretical motivations, the algorithms are illustrated in details, focusing on computational aspects and on the tuning of different parameters involved (as, for instance, the number of considered principal components, the scale of the bandwidth matrix in estimating the joint density $f_{d}$ and a ``mode cap'' parameter whose purpose is to prevent the springing of too much spurious modes). The study is completed with an analysis of real and synthetic datasets: a special attention is paid, in both clustering and discriminant framework, to mixtures presenting non-spherical clusters.

The paper is organised as follows: in Section~\ref{sec:theoretical} we introduce the mixture model and its factorization; Section~\ref{sec:classification} is devoted to illustrating the classification algorithms (clustering and discriminant) and to discussing the
parameters used; Section~\ref{sec:applications} collects the applications to simulated data and real cases; Section~\ref{sec:conclusion} gathers some final comments; finally, in \ref{sec:appendix} we sketch the proofs of the main theoretical results.

\section{Theoretical framework}
\label{sec:theoretical}

Let $X$ be a random element defined on the probability space $\left( \Omega ,\mathcal{F},\mathbb{P}\right) $, taking  values in the Hilbert space $\mathcal{L}_{\left[ 0,1\right] }^{2}$ endowed with the usual inner product $\left\langle \cdot, \cdot\right\rangle$ and associated norm $\|\cdot \|$. Denote by 
\begin{equation*}
\mu _{X}=\left\{ \mathbb{E}\left[ X\left( t\right) \right] ,t\in \left[ 0,1%
\right] \right\} ,\qquad \text{and}\qquad \Sigma \left[ \cdot \right] =%
\mathbb{E}\left[ \left\langle X-\mu _{X},\cdot \right\rangle \left( X-\mu
_{X}\right) \right] 
\end{equation*}%
the mean function and covariance operator of $X$ respectively. A measure of concentration of $X$ is given by the small--ball probability, briefly SmBP{} (see \citealp{fer:vie06} and reference therein), defined as
\begin{equation*} 
\varphi \left( x,\varepsilon \right) =\mathbb{P}\left( \left\Vert
X-x\right\Vert <\varepsilon \right) ,  \qquad x\in \mathcal{L}_{\left[ 0,1\right] }^{2},\ \varepsilon >0.
\end{equation*}
Suppose that $\Omega $ is
partitioned in $G$ (unknown and finite) sub-sets $\Omega _{g}$, and let $Y$ be a $\mathbb{N}$--valued random variable defined by 
\begin{equation*}
Y\left( \omega \right) =\sum_{g=1}^{G}g\, \mathbb{I}_{\Omega _{g}}(\omega
),\qquad \mathbb{P}\left( Y=g\right) =\pi _{g}>0,\qquad \sum_{g=1}^{G} \pi_{g}=1,
\end{equation*}%
(here $\mathbb{I}_{A}$ denotes the indicator of $A$) and consider the conditioned SmBP{} 
\begin{equation*}
\varphi \left( x,\varepsilon |g\right) =\mathbb{P}\left( \left\Vert
X-x\right\Vert <\varepsilon \ \left\vert \ Y=g\right. \right) ,\qquad
g=1,\ldots ,G,
\end{equation*}
that leads to the mixture
\begin{equation}\label{eq:SmBP_mixture}
	\varphi \left( x,\varepsilon \right) = \sum_{g=1}^{G} \pi_g \varphi \left( x,\varepsilon |g\right), \qquad x\in \mathcal{L}_{\left[ 0,1\right] }^{2},\ \varepsilon>0 .
\end{equation}
The latter expression is the starting point for approaching model--based classification problems: when $Y$ is a latent variable, we deal with an unsupervised classification problem focused on the left--hand side of \eqref{eq:SmBP_mixture}, see Section~\ref{sec:clustering_unsupervised_classification}; whereas, when $Y$ is observed, the model leads to the construction of a Bayesian classifier whose starting point is the right--hand side of \eqref{eq:SmBP_mixture}, see Section~\ref{sec:discriminant_supervised_classification}.
In both cases, instead of tackling it directly, we want to simplify it by exploiting an approximation result sketched below (for more details see \citealp{bon:goi15}). For the sake of simplicity, it is presented with respect to the process $X$; however, the same arguments can be applied to $ (X | Y = g) $ with $ g = 1, \ldots,  G$, provided a suitable change of notation is made.

Consider the Karhunen--Lo\`{e}ve expansion of $X$: denoting by $\left\{ \lambda _{j},\xi _{j}\right\} _{j=1}^{\infty }$ the decreasing to zero sequence of non--negative eigenvalues and their associated orthonormal eigenfunctions of $\Sigma $, it holds
\begin{equation*} 
X\left( t\right) =\mu _{X}\left( t\right) +\sum_{j=1}^{\infty }\theta
_{j}\xi _{j}\left( t\right) ,\qquad 0\leq t\leq 1,
\end{equation*}%
where $\theta _{j}=\left\langle X-\mu _{X},\xi _{j}\right\rangle $ are the
so--called principal components (PCs) of $X$ satisfying 
\begin{equation*}
\mathbb{E}\left[ \theta _{j}\right] =0,\qquad Var\left( \theta _{j}\right)
=\lambda _{j},\qquad \mathbb{E}\left[ \theta _{j}\theta _{j^{\prime }}\right]
=0,\qquad j\neq j^{\prime }.
\end{equation*}%
Without loss of generality, from now on suppose that $\mu _{X}=0$. Moreover, assume:
\begin{enumerate}[({A}.1)]
	\item\label{ass:A.1}
	the first $d$ PCs $\boldsymbol{\theta }=(\theta _{1},\ldots ,\theta _{d})^{\prime }$ admit a strictly positive and sufficiently smooth joint probability density $f_{d}$;
	
	\item
	$x$ is an element of $\mathcal{L}_{\left[ 0,1\right] }^{2}$ such that $\sup\{x_{j}^{2}/\lambda _{j}: {j\geq 1}\}<\infty$, with $x_j=\langle x, \xi_j\rangle$;

	\item
	there exists a positive constant $C$ (not depending on $d$) for which 
	\begin{equation*}
	\sup_{d\in \mathbb{N}}\sup_{i,j\in \{1,\ldots ,d\}}
	\frac{\sqrt{\lambda _{i}\lambda_{j}} }{\left\vert f_{d}(\boldsymbol{\vartheta })\right\vert}
	\left\vert \frac{\partial ^{2}f_{d}(\boldsymbol{\vartheta })}{\partial \vartheta _{i}\partial
		\vartheta _{j}}\right\vert  \leq C,\qquad \text{\textnormal{for
			any }}\boldsymbol{\vartheta }\in D,
	\end{equation*}%
	where $D=\left\{ \boldsymbol{\vartheta }\in \mathbb{R}^{d}:\sum_{j\leq
		d}\left( \vartheta _{j}-x_{j}\right) ^{2}\leq \rho ^{2}\right\} $ for some $%
	\rho \geq \varepsilon $;
	
	\item\label{ass:exp_decay}
	the spectrum of $\Sigma$ is rather concentrate: $\{\lambda _{j}\}_{j=1}^{\infty }$ decays to zero exponentially, that is
	\begin{equation} \label{eq:exp_decay}
		\lambda _{d}^{-1}\sum_{j\geq d+1}\lambda _{j}<C,\qquad \text{\textnormal{for any }}d\in \mathbb{N}.
	\end{equation}
\end{enumerate}

\begin{proposition}
	\label{pro:SMBP_approx} Under (A.1)--(A.4), as $\varepsilon $ tends to zero, it is possible to choose $d=d(\varepsilon )$ diverging to infinity so that: 
	\begin{equation}
	\varphi (x,\varepsilon )\sim f_{d}\left( x_{1},\dots ,x_{d}\right) \phi
	(d,\varepsilon).  \label{eq:SBP_factorization}
	\end{equation}

\end{proposition}
From a practical point of view, and as we will see in the sequel, the exponential decay required by Assumption~(A.\ref{ass:exp_decay}) is sufficient to reach the scopes of this paper. Nevertheless, to better appreciate the interpretation of factors in \eqref{eq:SBP_factorization} and, in particular, because the form of $\phi(d,\varepsilon )$ depends on them, we list below the forms of $\phi$ associated to the different eigenvalues decays. In particular:
\begin{itemize}
	\item 
	if $\{\lambda _{j}\}_{j=1}^{\infty }$ decays exponentially \eqref{eq:exp_decay}, then
	\[
	\phi (d,\varepsilon )=\exp
	\left\{ \frac{1}{2} d \left[ \log (2\pi e \varepsilon^2) - \log (d) +
	\delta(d, \alpha) \right] \right\},
	\]
	where $\delta(\cdot,\cdot)$ is such that $\lim_{\alpha\to\infty}
	\limsup_{s\to \infty} \delta(s, \alpha) = 0$ and $\alpha$ is a parameter
	chosen in such a way that $\lambda_d^{-1}\varepsilon^2\le \alpha^2$;
	
	\item  
	if $\{\lambda _{j}\}_{j=1}^{\infty }$ decays super--exponentially
	\begin{equation} \label{eq:super-exp_decay}
	\lambda_{d}^{-1} \sum_{j\geq d+1} \lambda_{j}\rightarrow	0,\qquad \textnormal{as}\qquad
	d\rightarrow \infty
	\end{equation}
	or, equivalently, $\lambda _{d+1}/\lambda _{d}\rightarrow 0 $ (as $d\to\infty $), then
	\[
	\phi(d,\varepsilon)=\exp	\left\{ \frac{1}{2} d \left[ \log (2\pi e \varepsilon^2) - \log (d) + \delta(d) 	\right] \right\},
	\]
	where $\delta(d)=o(1)$ as $d\to\infty$;
			
	\item 
	if $\{\lambda _{j}\}_{j=1}^{\infty }$ decays hyper--exponentially
	\begin{equation}\label{eq:hyper-exp_decay}
	d\left( \sum_{j\geq d+1}\lambda _{j} \right) \left( \sum_{j\leq d} \frac{1}{
		\lambda_j}\right) = o\left(1 \right),\qquad \textnormal{as }d \to \infty
	\end{equation}
	then
	\[
	\phi(d,\varepsilon)=\frac{\varepsilon^{d} \pi^{d/2}}{\Gamma \left( d/2+1\right)}.
	\]
\end{itemize}

The fact that in the hyper--exponential case, $\phi(d,\varepsilon)$ is the volume of a $d$--dimensional ball with radius $\varepsilon$, justifies to interpret, in Equation~\eqref{eq:SBP_factorization}, $\phi(d,\varepsilon)$ as a $d$--dimensional volume parameter, whilst $f_d$, being the only factor depending on $x\in \mathcal{L}_{\left[ 0,1\right] }^{2}$, as a surrogate of the density of the Hilbert process.
\begin{remark} 
	Note that \eqref{eq:hyper-exp_decay} $ \Rightarrow$ \eqref{eq:super-exp_decay} $\Rightarrow$ \eqref{eq:exp_decay}, whereas the vice versa does not hold. For instance, for any $\alpha >1$ and $\beta >0$,  $\lambda_{d} = \exp\left\{-\beta d \right\}$ decays exponentially but not super--exponentially, $\lambda_{d} = \exp\left\{-\beta d \ln\left( \ln\left(d \right)\right) \right\}$ decays super--exponentially but not hyper--exponentially while $\lambda_{d} = \exp\left\{-\beta d^\alpha \right\}$ decays hyper--exponentially.
\end{remark}


\begin{remark} 
	Although the results are exposed using the Karhunen--Lo\`{e}ve (or PCA) basis, they hold for any orthonormal basis ordered according to the decreasing values of the variances of the projections, provided that they decay sufficiently fast (see \citealp{bon:goi15}).
	Note that the variances obtained when one uses the PCA basis present, by construction, the faster decay: in this sense the choice of this basis can be considered optimal.
\end{remark}

\section{Classification}
\label{sec:classification}

This section is devoted to defining classification procedures in a functional framework. The idea is to exploit the asymptotic factorization results provided by Proposition~\ref{pro:SMBP_approx}.
It is divided in two subsections: the first one illustrates a clustering algorithm, whereas the second one deals with a discriminant analysis procedure.

\subsection{Unsupervised classification}
\label{sec:clustering_unsupervised_classification}

Consider a sample $\{ (X_1,Y_1), \ldots, (X_n,Y_n) \}$ drawn from $(X,Y)$ defined as in Section \ref{sec:theoretical}, where $X$'s are observed while the group variables $Y$'s are latent. Our aim is to determine the range of $Y$ (i.e.\@ $G$) and, for each observed $X_i$ the membership group (that is the value of $Y_i$).
If the distribution of $(X|Y=g)$ is specified then a full parametric approach applies; this has been done, for example, in \cite{jac:pre14} where the authors used a maximum likelihood and expectation maximization approach to identify the distribution parameters of a Gaussian mixture assumed for $f_d$.
Instead, if no information is available, a distribution free model could be used. In this latter view, consider the SmBP{} mixture \eqref{eq:SmBP_mixture} and apply Proposition~\ref{pro:SMBP_approx} to its left--hand side to obtain:
\[
	\sum_{g=1}^{G} \pi_g \varphi \left( x,\varepsilon |g\right) = \varphi \left( x,\varepsilon \right) \sim f_{d}(x_{1},\dots ,x_{d})\phi (d,\varepsilon ) , \qquad x\in \mathcal{L}_{\left[ 0,1\right] }^{2},\ \varepsilon\to 0 .
\]
Such expression highlights how the surrogate density $f_d$ carries the information on the mixture and, at the same time, endorses a ``density oriented'' clustering approach on $f_d$ as a fruitful tool in detecting the latent structure; hence, from now on, we assume that there exists a positive integer $d^\star$ such that $f_d$ is $G$--modal for any $d\ge d^\star$. Therefore, our scope is to identify the groups by a ``locally high (surrogate) density regions'' principle: the clustering algorithm computes the estimates $\widehat{m}_{d,g}$ of the modes $m_{d,g}$, that is the local maxima of $f_d$, whose number $\widehat{G}$ estimates $G$; for each $g$, it finds the largest connected upper--surface containing only $\widehat{m}_{d,g}$, and hence it assigns group labels to each observation consistently with its proximity to these sets.
The algorithm procedure is illustrated below:
\begin{enumerate}[Step 1.]
	\item
	Estimate the covariance operator and the eigenelements.
	\item
	Fixed $d$, compute $\widehat{f}_{d,n}$ (an estimation of the joint distribution density $f_{d}$).
	\item
	Look for the local maxima $\widehat{m}_{d,g}$ of $\widehat{f}_{d,n}$, $g=1,\ldots,\widehat{G}$ over a grid.
	\item
	\emph{Finding Prototypes}: for each $g$ in $\{1,\ldots, \widehat{G}\}$, the $g$-th ``prototypes'' group is formed by those $X_i$ whose estimated PCs $(\widehat{\theta}_{1,i}, \ldots, \widehat{\theta}_{d,i})$ belong to the largest connected upper--surface of $\widehat{f}_{d,n}$ that	contains only the maximum $\widehat{m}_{d,g}$. In other words, for such individual, $\widehat{Y}_i=g$.
	\item
	Assign each unlabelled $X_{i}$ to a group by means of a k--NN procedure (with $k=1$).
	
\end{enumerate}

As a by--product of such algorithm, it is possible to define a center for each cluster through the ``$d$--dimensional modal curves'' built from $\widehat{m}_{d,g}$ and defined as follows:
\[
	\widehat{X}^m_g(t) = \sum_{j=1}^d \widehat{m}_{d,g}^{(j)} \widehat{\xi}_j(t)
\]
where $\widehat{m}_{d,g}^{(j)}$ is the $j$--th term of $\widehat{m}_{d,g}$ and $\widehat{\xi}_j(t)$ are the empirical versions of $\xi_j(t)$.
\\%
In the remaining part of this section, some theoretical and practical aspects of the algorithm will be discussed.

\subsubsection{Surrogate density estimation}
\label{sec:surrogate_density_estimation_and_H_choice}

In order to estimate $f_{d}$, we consider the classical multivariate kernel density estimator:
\begin{equation*} 
\widehat{f}_{d,n}\left( \widehat{\Pi }_{d} x \right)  = \frac{1}{n}\sum_{i=1}^{n}K_{H} \left( \left\Vert \widehat{\Pi }_{d}\left( X_{i}-x\right) \right\Vert \right) ,\qquad \widehat{\Pi }_{d} x \in \mathbb{R}^{d},
\end{equation*}
where $K_{H}\left( \mathbf{u}\right) =\det \left(H\right)^{-1/2}K\left( H^{-1/2}\mathbf{u}\right) $, $K$ is a kernel function, $H$ is a symmetric semi-definite positive $d\times d $ matrix and, $\widehat{\Pi }_{d}\left(\cdot \right) = \sum_{j=1}^d \widehat{\xi}_j \left\langle \widehat{\xi}_j, \cdot \right\rangle $ is the projection operator over the subspace spanned by $\left\{ \widehat{\xi}_1, \ldots, \widehat{\xi}_d\right\}$, i.e. the first $d$ eigenfunctions of $\widehat{\Sigma}_n$ (the sample version of $\Sigma$) so that the kernel argument depends on the estimated PCA semi--metric \cite[Section~8.2]{fer:vie06}.
It is worth noticing that in estimating $f_d$ the use of the estimated projector $\widehat{\Pi}_d$ instead of $\Pi_d$ introduces a non--standard source of noise that might modify  the consistency properties and, hence, should be considered with care. In fact, from a theoretical point of view, one may wonder if such estimator is consistent for $f_d$ and, when this is the case, if it attains the same rate of convergence that holds when $\Pi_d$ is known. 
A positive answer was provided in \cite{bon:goi15}; in particular, consider the special case $H_{n}=h_{n}^{2}I$, and suppose that:
\begin{enumerate}[({B}.1)]
	\item\label{ass:B1}
	 the density $f_{d}\left( x\right) $ is positive and $p$ times
	differentiable at $x \in\mathbb{R}^d$;
	
	\item 
	 $h_{n}$ is such that $h_{n}\rightarrow 0$ and $nh_{n}^{d}/\log n\rightarrow \infty$ as $n\rightarrow \infty $;
	
	\item 
	 the kernel $K$ is a Lipschitz, bounded, integrable density function with compact support $\left[ 0,1\right] $;
	
	\item\label{ass:B4} the process $X$ is bounded.
\end{enumerate}
Then, the following result holds:
\begin{proposition} \label{prop:rate_convergence}
	Assume (B.\ref{ass:B1})--(B.\ref{ass:B4}) with $p>\left( 3d+2\right) /2$ and consider the optimal bandwidth 
	\begin{equation}\label{eq:optimal bandwidth}  
		c_{1}n^{-\frac{1}{2p+d}}\leq h_{n}\leq c_{2}n^{-\frac{1}{2p+d}}
	\end{equation}
	where $c_{1}$ and $c_{2}$ are two positive constants. Thus, as $n$ goes to infinity,
	\begin{equation*}
	\mathbb{E}\left[ f_{d}\left( x\right) -\widehat{f}_{d,n}\left( x \right) \right]
	^{2}=O\left( n^{-2p/\left( 2p+d\right) }\right),
	\end{equation*}
	uniformly in $\mathbb{R}^d$.
\end{proposition}

From a practical point of view, the bandwidth selection is an important task: our choice is to consider a diagonal bandwidth matrix (see \citealp{duo:haz05} for a heuristic justification) whose non--null entries are the univariate bandwidth provided by \citet[p.48]{sil86}. Anyway, it is clear that the larger is $|H|$, the ``smoother'' is $\widehat{f}_{d,n}$, the smaller is the number of modes and hence the number of groups. In other words, different choices for the bandwidth may be considered in order to catch different phenomenon scales; this is done by applying to $H$ a scale factor $\delta >0$, whose optimality (in a sense to be specified) is discussed below.

\subsubsection{Prototypes identification and Modes}

The identification of prototypes is the core of the algorithm: it rests on the largest connected upper--level sets of $\widehat{f}_{d,n}$ related to each mode $\widehat{m}_{d,g}$. The latter plays just an instrumental part in identifying prototypes.
More in details, we use the graphical visualization system of R software: $X_{i}$ is assigned to the $g$--th prototype group if its estimated PCs belong to the $g$--th upper--level sets, by using the algorithm described in \cite{liu:che:mai:lut10} and available in the R--package \emph{ptinpoly}. 
From a practical point of view, a problem concerns spurious modes caused by the choice of $\delta$ and sampling variability. To tackle this issue, we select only those modes $\widehat{m}_{d,g}$ for which $\widehat{f}_d(\widehat{m}_{d,g})$ is maximum over the parallelepiped $\widehat{m}_{d,g} \pm [0, r h_1]\times \ldots\times [0,rh_d]$, where $h_j$ is the resolution along the $j$--th direction of the grid used to estimate the density, while $r$ is a positive integer, playing the counterpart of a tolerance coefficient.

It is worth recalling that, in the multivariate literature, alternative techniques in detecting high density regions can be found (see for instance, \citealp{azz:tor07}, \citealp{sag79}, \citealp{rin:sin:nug:was12} and references therein).
	
To conclude, we provide a glimpse of the theoretical aspects about mode estimation. For a fixed $d$, consistency properties for estimated modes have been considered, for instance, in \cite{abr:bia:cad03}, \cite{che:gen:was15} and \cite{sag79}. Nevertheless, such theoretical issues are little studied in the infinite dimensional framework, since the mathematical concept of the density is still under--developed; some attempts can be found in \cite{gas:hal:pre98} and, more recently, in \cite{bon:goi15}, \cite{dab:fer:vie07} and \cite{del:hal10}.

\subsubsection{Tuning parameters}\label{sec:parameters_tuning}

Given a data set, different values of $d$, $r$ and $\delta$  may lead to different cluster results: thus, in many situations, a criterion to choose them is opportune. In literature it is acknowledged that, for the most clustering algorithms, there is not a selection ``golden rule'' of the parameters. In practice, a series of trials and repetitions are performed to tune the parameters (see \citealp{xu:wun05} and references therein). In this view, automatic selection rules can be used only as support in decisional steps.
\newline
In view of Proposition~\ref{pro:SMBP_approx}, parameter $d$ should be large enough to guarantee a good approximation for the SmBP{} \eqref{eq:SBP_factorization}, but small enough to avoid the well--known ``curse of dimensionality'' in estimating non--parametrically $f_{d}$.
A good compromise is to choose $d$ so that the Fraction of Explained Variance, $FEV(d) = {\sum_{j\le d}\lambda_j}/{\sum_{j\ge 1}\lambda_j}$, is larger than a suitable constant.
In practice, $FEV$ is estimated from eigenvalues of $\widehat{\Sigma}$.

For what concerns a choice for $(r,\delta)$, and to provide some insights on the quality of clustering solutions, we exploit both external and internal criteria although they are not universal and effective tests for an optimal choice. In particular, we implement: the purity index (based on some prespecified structure, which is the reflection of prior information on the data), the ``Cali\~{n}ski and Harabasz'', or briefly ``CH'', index (that does not depend on external information) and, when feasible, a combination of the two indices. Thus, accordingly to the nature of data, the idea is to look at those $(r,\delta)$ which furnish the best value for these validation criteria and, as a consequence, suggesting the number of clusters $G$. In the remaining part of this section, we summarise these two criteria; more details can be found in \cite{xu:wun05} and references therein.
\newline
The purity index measures how close a clustering is to an available pre--specified class structure and, more precisely, the extent to which each cluster consists of objects from a single class.
In particular, for each cluster, consider the class distribution of the data; i.e. for class $j$ compute $p_{gj}$, the frequency a member of cluster $g$ belongs to class $j$ as $p_{gj}=\pi_{gj}/\pi_{g}$, where $\pi_g$ is the proportion of objects in cluster $g$, and $\pi_{gj}$ is the proportion of objects of class $j$ in cluster $g$. Hence, for each cluster $g$, purity is calculated as
\begin{equation*}
p_g = p_g(r,\delta) = \max \{p_{gj} : j=1,\ldots,L\},
\end{equation*}
where $L$ is the number of pre--specified classes, whilst the total purity is the sum of the cluster purities weighted by the size of each cluster $p =\sum_{g=1}^{G} p_g \pi_g$. Clearly, $p$ ranges in $[0,1]$ with $p=0$ meaning maximum separation and $p=1$ maximum cohesion.
\newline
Among the internal validation indices, the $CH$ index is well--known and often achieves the best performance (see \citealp{dub93}). It is defined as
\[
CH = CH(\delta, r) = \left\{
\begin{array}{ll}
\frac{Tr(S_B)}{K-1} \Big/ \frac{Tr(S_W)}{n-K}, & K>1,
\\
0, & K=1.
\end{array}
\right.
\]
where $N$ is the sample size, $K$ is the number of clusters obtained by choosing the couple $(\delta, r)$, and $Tr(S_B)$ and $Tr(S_W)$ are the traces of the estimated between and within covariance matrices, respectively. The couples $(\delta, r)$ that maximize $CH$ are selected as optimal, and the number of clusters $K$ is consequently obtained.
\newline
It is worth noticing that purity is not affected by the geometry of the point clouds, whereas $CH$ is. In particular, since $CH$ definition is based on variances, it is expected that $CH$ provides good results whenever the clusters are elliptical point clouds and not otherwise.

\subsection{Supervised classification}
\label{sec:discriminant_supervised_classification}

In discriminant analysis, differently from clustering, the presence of $G$ distinct groups is established and modelled by the observed variable $Y$: the aim is to label each new incoming observation according to this known group structure. To do this, a typical approach is to use a \emph{Bayes classification rule}: given an observation $x$, one assigns it to the class $\gamma (x)\in \{1,\ldots ,G\}$ to which corresponds the highest {\it a posteriori} probability $\mathbb{P}(Y=\gamma (x)|X=x)$:
\[
	\gamma (x)=\arg \max_{g=1,\ldots ,G}\mathbb{P}(Y=g|X=x).
\]
Equivalently, $\gamma (x)$ is the index $g^{\prime }$ in $\{1,\ldots ,G\}$
such that 
\begin{equation}\label{eq:posteriors_ratio}
	\frac{\mathbb{P}(Y=g^{\prime }|X=x)}{\mathbb{P}(Y=g|X=x)}>1,\qquad \text{\textnormal{for any }}g=1,\ldots ,G\text{ \ \textnormal{and }}g\neq g^{\prime }.  
\end{equation}
If a probability density of $X$ in the $g$--th group $f(x|g)$ were known (with $f(x|g)>0$), thanks to the Bayes formula, Equation~\eqref{eq:posteriors_ratio} would simplify as follows:
\[
	\frac{\pi _{g^{\prime }}f(x|g^{\prime })}{\pi _{g}f(x|g)}>1,\qquad \text{\textnormal{for any }}g\neq g^{\prime },
\]
and, consequently, the classification rule would become:
\[
	\gamma (x)=\arg \max_{g=1,\ldots ,G}\pi _{g}f(x|g).
\]
It is clear that such arguments do not apply straightforwardly in functional settings without further assumptions on the probability measures. A possible way to tackle the problem is to consider the following classification rule: assign a new functional observation $x$ to the $g$--th group for which, as $\varepsilon $ tends to $0$,
\begin{equation}\label{eq:discr_rule_initial}
	\frac{\mathbb{P}(Y=g\ |\ \left\Vert X-x\right\Vert <\varepsilon )}{\mathbb{P}(Y=g^{\prime }\ |\ \left\Vert X-x\right\Vert <\varepsilon )}>1,\qquad \text{\textnormal{for any }}g^{\prime }\neq g. 
\end{equation}
At a glance, it is evident that it is hard to use in practice. Anyway, thanks to the Bayes formula, the ratio in \eqref{eq:discr_rule_initial} becomes
\begin{equation*} 
\frac{\pi _{g}\mathbb{\varphi }\left( x,\varepsilon |g\right) }{\pi	_{g^{\prime }}\mathbb{\varphi }\left( x,\varepsilon |g^{\prime }\right) }. 
\end{equation*}
Whenever assumptions of Proposition~\ref{pro:SMBP_approx} hold for each $(X|Y=g)$ with $g=1,\ldots, G$, then the above ratio reduces to
\[
	\frac{\pi _{g}f_{d_{g}}\left( x|g\right) \phi (d_{g},\varepsilon )}{\pi_{g^{\prime }}f_{d_{g^{\prime }}}\left( x|g^{\prime }\right) \phi (d_{g^{\prime }},\varepsilon )},\qquad \text{\textnormal{as }}\varepsilon\rightarrow 0,
\]
where $f_{d_{g}}\left( x|g\right)$ is the joint density of the first $d_g$ PCs computed using the (conditional) covariance operator $\Sigma_g$ of the group $g$.
Clearly, the asymptotic behaviour of the ratio is related to the trade--off between the volume parameters $\phi (\cdot,\varepsilon )$ and the probability densities evaluated at possibly different dimensions $d_{g}$ and $d_{g^{\prime }}$.

The classification rule may be further simplified if additional assumptions are imposed on the mixture process $X$, since the spectrum decay of $\Sigma$ controls the one of each $\Sigma_g $ (the conditional covariance operator corresponding to the $g$--th group), the starting point to build $f_{d_g}$.
In particular, consider the variance decomposition $\Sigma = B + W $, where $B$ and $W=\sum_{g\in G} \pi_g \Sigma_g$ represent the between and within covariance operator respectively. A straight application of the Courant--Fischer--Weyl min--max principle for linear operators leads to 
\begin{align*}
\lambda_k(M_1) \le \lambda_k (M_1+M_2),
\end{align*}
where $\{\lambda_{j}(M_i)\}$ denote the eigenvalues of the linear operator $M_i$ (in decreasing order).
The latter inequality ensures that the eigenvalues of $B$, $W$ and $\Sigma_g$ ($g=1,\ldots,G$) have a decay fast at least as the one of $\Sigma$. In other words, since the eigenvalues' decay rate is a measure of how much $X$ is concentrated in the space, the process $(X|Y=g)$ in each sub--population must be ``concentrated'' at least as much as $X$.
As a consequence, if the spectrum of $\Sigma$ decays exponentially (according to \eqref{eq:exp_decay}), then $d_g$ can be chosen equal to $d$ for any $g=1,\ldots ,G$, $\phi (d,\varepsilon )$ simplifies, and one can write the classification rule \eqref{eq:discr_rule_initial} similarly to the multivariate case, by replacing a probability density with a surrogate version: assign a new functional observation $x$ to the $g$--th group for which, as $\varepsilon $ tends to $0$,
\[
	\frac{\pi _{g}f_{d}\left( x|g\right) }{\pi _{g^{\prime }}f_{d}\left(x|g^{\prime }\right) }>1,\qquad \text{\textnormal{for any }}g^{\prime }\in \{1,\ldots ,G\},\ g^{\prime }\neq g,
\]
or, equivalently, as $d$ tends to $+\infty $,
\begin{equation}\label{eq:discr_rule_final}
	\gamma (x,d)=\arg \max_{g=1,\ldots ,G}\pi _{g}f_{d}(x|g).
\end{equation}
Operatively, if one could specify the conditional densities $f_{d}(x|g)$, a full parametric approach would be possible. Although $\gamma (x,d)$ still depends on $\varepsilon $ by means of $d$ (see Proposition~\ref{pro:SMBP_approx}), it is not so restrictive to assume that 
\begin{enumerate}[({A}.1)]
	\setcounter{enumi}{4}
	\item \label{ass:A.5}
	$\gamma (x,d)$ is constant as $d$ goes to infinity; i.e.\@ there exists a positive integer $d^{\ast }$ such that $\gamma (x,d)=\gamma (x,d^{\ast })$ for any $d\geq d^{\ast }$.
	
\end{enumerate}
This assumption holds, at least, in the case of a finite dimensional process.

At this point, a comparison of our approach with the one introduced in \cite{jam:has01}  is interesting, and one can trace some parallelism. Indeed, in both approaches, a dimensionality reduction step based on projection onto a finite vector subspace generated by a previously chosen basis is implemented, and so the classification rule involves the conditional joint densities of the projection coefficients. Moreover, if $d_{g}=d$ for all $g$, and one assumes an underlying Gaussian mixture model, both approaches lead to the same classifier: the present section theoretically justifies the use of \eqref{eq:discr_rule_final} in a finite dimensional subspace.
However, if the eigenvalues of $\Sigma$ decay slowly, we cannot ensure that $d_{g}$ is the same varying $g$, and the volume terms $\phi(d_{g},\varepsilon )$ cannot be neglected in the classification rule. Consequently, the approach based on a pure projective method could be unfruitful.

The illustrated method can be framed by the full parametric discrimination described in \cite{jam:has01} and the full nonparametric one proposed by \cite{fer:vie03}, where the {\it a posteriori} probability is estimated directly by a kernel regression approach.

\subsubsection{Estimate classifier}

Once $d$ is chosen, since we want to work in a distribution--free context, densities $f_{d}(x|g)$ have to be estimated. 
Consider a sample $\left\{ \left( X_{1},Y_{1}\right) ,\dots ,\left(X_{n},Y_{n}\right) \right\} $ from $(X,Y)$, with $d_{g}=d$ for
each $g=1,\dots ,G$; a kernel density based estimator for %
\eqref{eq:discr_rule_final} is given by:%
\begin{equation}\label{eq:discr_rule_estimate}
\widehat{\gamma }_{n}(x,d)=\arg \max_{g=1,\ldots ,G}\frac{\widehat{\pi }_{g}}{n_{g}}\sum_{i=1}^{n}\mathbb{I}_{\left\{ Y_{i}=g\right\} }\ K_{H_{g}}\left(\left\Vert \widehat{\Pi }_{g,d}\left( X_{i}-x\right) \right\Vert \right)  
\end{equation}%
where $n_{g}=\sum_{i=1}^{n}\mathbb{I}_{\left\{ Y_{i}=g\right\} }$ is the
number of observations coming from group $g$, $\widehat{\pi }_{g}=n_{g}/n$
estimates the mixture coefficient $\pi _{g}$, $K_{H_{g}}\left( u\right)
=\det \left( H_{g}\right) ^{-1/2}K\left( H_{g}^{-1/2}u\right) $, with $K$ a
kernel function, the bandwidth matrix $H_{g}$ is  $d\times d$ and symmetric
semi-definite positive, and finally $\widehat{\Pi }_{g,d}$ is the projection
operator over the subspace spanned by the first $d$ eigenfunctions of the
sample covariance operator $\widehat{\Sigma }_{g}$ for the group $g$. 

Note that some simplifications may occur in \eqref{eq:discr_rule_estimate}.
For instance, if the groups are balanced, that is $\pi _{g}=1/G$ for each $g$%
, then one can drop $\widehat{\pi }_{g}$ and $n_{g}$. Another example
concerns the homoscedasticity case (i.e.~$\Sigma _{g}$ is the same for each $%
g$), where $\widehat{\Pi }_{g,d}=\widehat{\Pi }_{d}$ is the projector over
the space spanned by the first $d$ eigenfunctions of the within (or pooled)
covariance operator $\widehat{W}=\sum_{g=1}^{G}\widehat{\pi}_g\widehat{\Sigma }_{g}$.

For what concerns the choice of $d$ and $H_{g}$, one can refer to the
discussion in Section~\ref{sec:parameters_tuning}. It is scarcely
necessary to observe that, in the discriminant context, coefficients $r$ and 
$\delta $ are not necessary, because we have a specific bandwidth matrix for
each group and the main goal is not the estimation of a mode.

To complete the analysis, we study the asymptotic properties of the
classifier $\widehat{\gamma }_{n}$ defined in \eqref{eq:discr_rule_estimate}. 
In particular, we consider the Bayes probability of error
$$
L^{\star }=\min_{\gamma }\mathbb{P}\left( \gamma \left( X\right)
\not=Y\right)
$$
and the conditional probability of error
$$
L_{n}=\mathbb{P}\left( \widehat{\gamma }_{n}\left( X,d\right) \not=Y\ |\
\left\{ \left( X_{1},Y_{1}\right) ,\dots ,\left( X_{n},Y_{n}\right) \right\}
\right)
$$
and we study how $L_{n}$ behaves when $n$ tends to infinity.
In particular, convergence of $L_{n}$ to the Bayes error probability $L^{\star }$ is stated
in the following proposition, which is a direct consequence of the results by \cite{dev81}, Section 5.

\begin{proposition} 
	Take $H_g=h_g I$. Under assumptions ~(A.\ref{ass:A.1})--(A.\ref{ass:A.5}), $L_{n}$ converges to $L^{\star }$ in probability, as $n$ tends to $+\infty$.
\end{proposition}

\section{Applications to synthetic and real data}
\label{sec:applications}

This section concerns the simulations and applications of the previously described methods: the first two subsections are dedicated to clustering (\ref{sec:clustering_simulations} considers an experiment under a controlled set-up, whereas in \ref{sec:clustering_real_data} we apply clustering to a real dataset), and the last one (Section~\ref{sec:discriminant_simulation_real_data}) is dedicated to discriminant analysis.

\subsection{Clustering: simulation examples and comparison with competitors}
\label{sec:clustering_simulations}

In the following, a simulation study provides a quantitative comparison of the presented algorithm versus competitors. 
Although the methods are unsupervised, we evaluate their ability in detecting the underlying group structure by measuring a misclassification error, as if it was a supervised exercise. Besides, by construction the SmBP clustering provides an estimate of the number of clusters that must be studied as well, it being a source of noise.
As pointed out in Section~\ref{sec:parameters_tuning}, both misclassification error and number of detected clusters depend on the choice of parameters: keeping this in mind, the simulation exercise is coherently calibrate.

In order to generate the dataset, we use the functional basis expansion: 
\begin{equation*} 
	X_{i}^{(g)}(t)=\sum_{l=0}^{L}\sqrt{\beta_{l}}\, \tau_{i,l}^{(g)}\psi_{l}(t),\qquad t\in \lbrack 0,1],\ i=1,\ldots ,n_g\ \text{ and }\ g=1,\ldots,G,
\end{equation*}%
where $\beta_{l}=0.7\times 3^{-l}$ ($l=1,\ldots, L=150$) and $\psi _{l}(t)$ is the $l$-th element of the Fourier basis 
\begin{equation*}
	\psi _{l}(t)=\left\{ 
	\begin{array}{l}
	\sqrt{2}\sin (2\pi mt-\pi ),\ l=2m-1; \\ 
	\sqrt{2}\cos (2\pi mt-\pi ),\ l=2m.%
	\end{array}%
	\right.
\end{equation*}
The mixture is controlled by means of $\tau^{(g)}$'s. Here, we deal with $G=2$ and, to avoid spherical shaped groups, uncorrelated but dependent coefficients  $(\tau_{i,l}^{(g)})_{l=1}^{L}$ are generated as follow:
\begin{equation*}
\left\{
\begin{array}{rcll}
\tau_{i,1}^{(g)} & = & \sin(\vartheta_{i})\cos(\frac{\pi}{2} \mathbb{I}_{\{g=2\}}) + \sigma \epsilon_{i,1} &
\\
\tau_{i,2}^{(g)} & = & \sin(\vartheta_{i})\sin(\frac{\pi}{2} \mathbb{I}_{\{g=2\}}) +\sigma \epsilon_{i,2} &
\\
\tau_{i,3}^{(g)} & = & \cos(\vartheta_{i}) + (-k)^g +\sigma \epsilon_{i,3} &
\\
\tau_{i,l}^{(g)} & = & \sqrt{0.1}\epsilon_{i,l}, & 4\le l\le L
\end{array}
\right.
\end{equation*}
with $(\vartheta_i)$ i.i.d.\@ as a {\it Beta(5,5)} scaled on $[-\pi,\pi]$ and $(\epsilon_{i,l})_{l=1}^L\stackrel{i.i.d.}{\sim} {\cal N}(0,1)$. In other words, $(\tau_{i,l}^{(g)})_{l=1}^3$ are the Cartesian coordinates of the spherical ones $(1, \theta_{i}^{(g)},\frac{\pi}{2} \mathbb{I}_{\{g=2\}})_{l=1}^3$ plus  a vertical translation $(-k)^g$ and a Gaussian noise $\epsilon$ (randomness is confined in the polar angle $\vartheta$ and in the noise $\epsilon$). In particular, limited to the first three components, we deal with two noised semi--circumferences laying on orthogonal planes, with unitary radii, whose centers are $(0,0,\pm k)$ and chosen so that the clouds of points of $(\tau_{i,l}^{(g)})_{l=1}^3$ look like two interlocked horseshoes.
A reasonable range for $k$ is $(0,1)$: outside this range, the two un--noised groups can be separated by means of a plane, a structure easily identifiable.
Concerning $\sigma$, one can choose $(0, k/3)$ to avoid that groups overlap too much due to noise variability. 
\\
With such choices, we are concentrating the process along three orthogonal directions so that the PCs tend to replicate the $\tau$'s structure. Moreover, this setting ensures that Proposition~\ref{pro:SMBP_approx} applies. In fact, the eigenvalues (of $\Sigma$) decay faster or at least equally to $\{\beta _{l}Var(\tau_l)\}_{l=1}^L$. Due to boundedness of $Var(\tau_l)$, it inherits the same decay type of $\{\beta _{l}\}_{l=1}^L$ that is exponentially \eqref{eq:exp_decay} with $C=1/3$.

\begin{figure}[!htb]
	\begin{center}
		\includegraphics[clip=true, trim = 14cm 2cm 14cm 2cm, width = .32\textwidth, height =6cm]{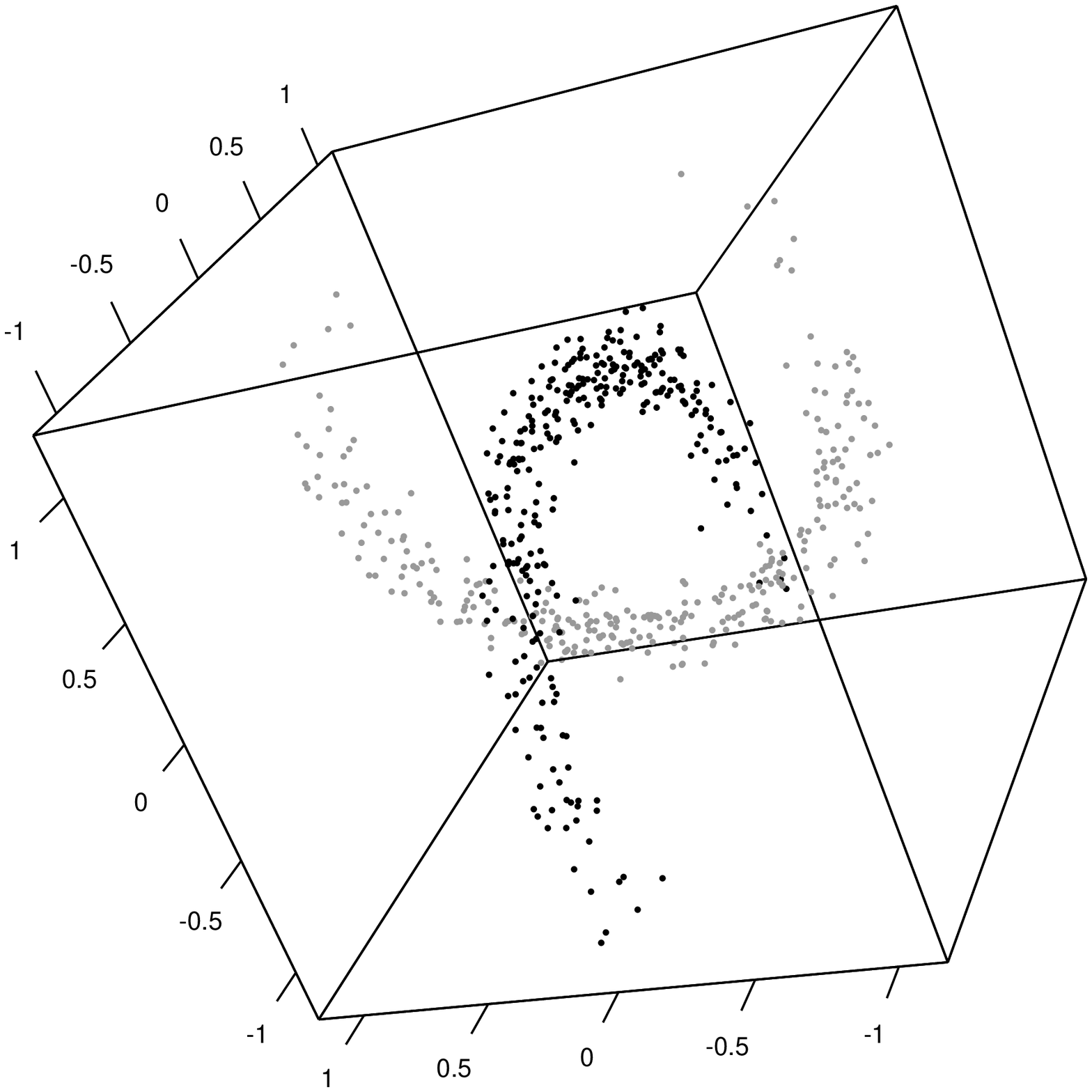}
		\includegraphics[width = .32\textwidth, height =5.5cm]{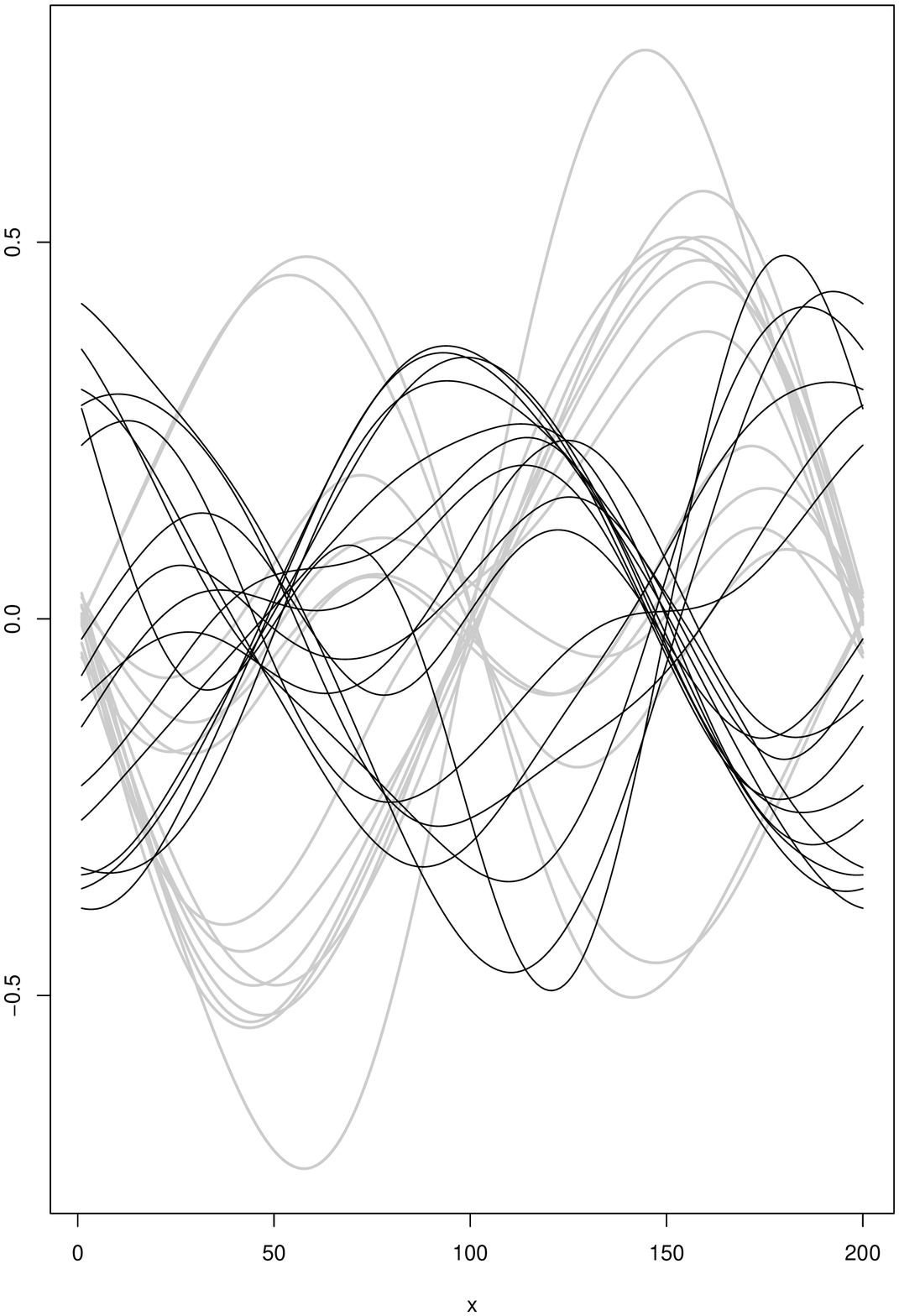}
		\includegraphics[clip=true, trim = 14cm 2cm 14cm 2cm, width = .32\textwidth, height =6cm]{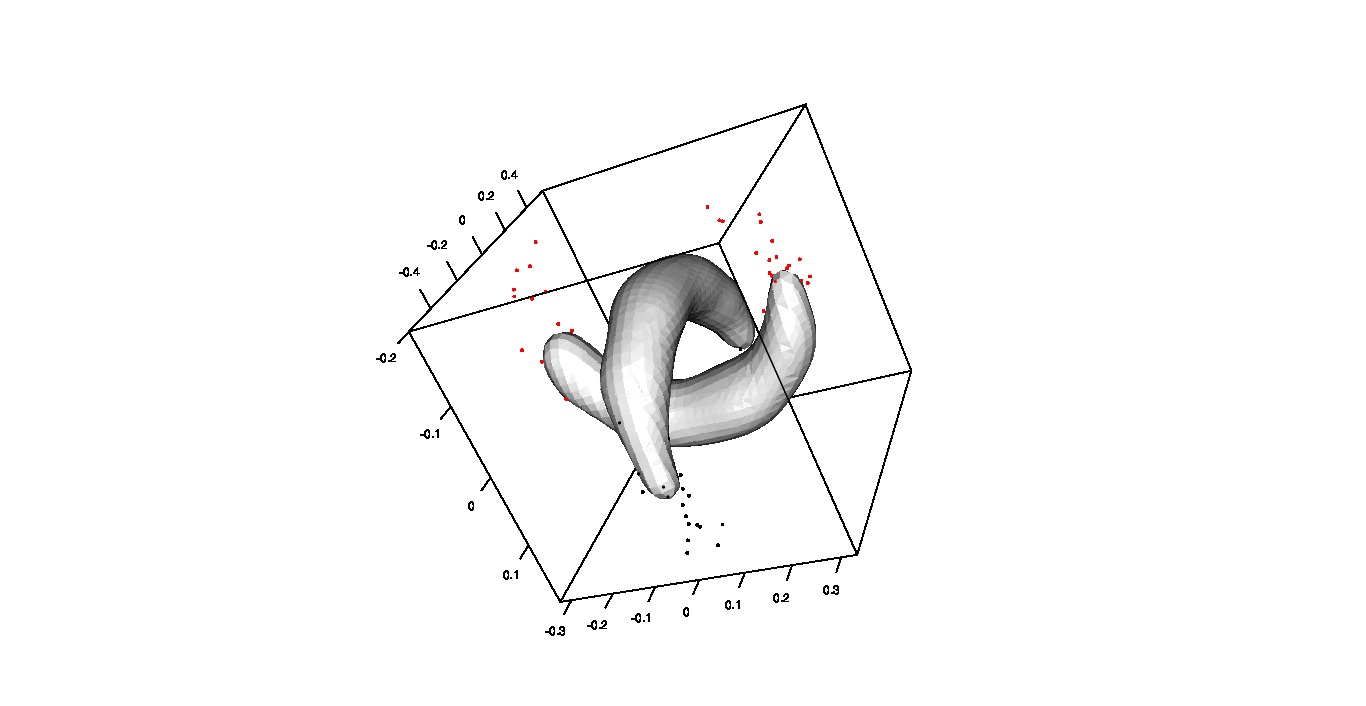}
	\end{center}
	\caption{Left to right: simulated coefficients $(\tau_{i,l}^{(k)})_{l=1}^3$, a sample of simulated curves and upper level sets associated to the estimated modes in the factor space.}
	\label{fig:bananas}
\end{figure}

In our simulations, we consider $n_1=n_2=300$, $\sigma = \sqrt{0.005}$ and $k=0.5$. This setting leads to have $FEV(3)$ always greater than 99\%, that suggests us to fix $d=3$. Curves are generated over a grid of 100 equispaced points on $[0,1]$. For the sake of illustration, Figure~\ref{fig:bananas} depicts the scatter plot of an observed set of $(\tau_{i,l})_{l=1}^3$, a selection of the corresponding curves and the prototype regions obtained with our algorithm when $\delta = 2$ and $r = 5$ (that produced $\widehat{G}=2$).

We generate $400$ Monte Carlo samples according to the above setting. To each replication, we apply the SmBP{} clustering that returns the corresponding estimated number of clusters $\widehat{G}$ and the misclassification error. According to $FEV$ criterion we set $d=3$, and we explore the behaviour of the algorithm when $\delta = 0.6, 1, 1.4$ and $r = 1, 5, 10$. The following competitors are considered:
\begin{itemize}
	\item[(KM)]
	the functional $k$--means clustering (see \citealp{feb:ovi12}) with $G=2$;
	
	\item[(GM)]
	the EM clustering method based on a mixture model of Gaussian components applied to the first three PCs. We consider both $G=2$ and $G$ estimated by means of a Bayesian Information Criterion (BIC). The algorithm is coded in package \textit{Rmixmod} (see \citealp{bie:cel:gov:lan06}).
\end{itemize}
Table~\ref{tab:bananas_misclas_err} collects the main results. For SmBP{} clustering, we report the mean and the standard deviation of misclassification errors and the $0.5$, $0.75$, $0.9$ order quantiles of $\widehat{G}$ varying $\delta$ and $r$; the same results are provided for (GM) combined with BIC, and only the misclassification error whenever $G$ is fixed.
Such results show that there exists an optimal configuration of the parameters $\delta=1.4$ and $r=10$ for the SmBP clustering for which the two clusters are correctly recognized at least in 90\% of the cases, with an average misclassification error equal to 8.8\%. It can be noted that the parametric method (GM) produces good results whenever $G$ is fixed, but gets worse when $G$ has to be estimated, since BIC overestimates the number of clusters.

\begin{table}[!htb]
\begin{center}
\begin{tabular}{lccccccc} 
	\hline
	Algorithm & \multicolumn{2}{c}{Parameters} & \multicolumn{2}{c}{
		Miscl. Error} & \multicolumn{3}{c}{$\widehat{G}$} \\
	\hline\hline
	& $\delta $ & $r$  & Mean & St.~Dev. & $q_{0.5}$ & $q_{0.75}$ & $q_{0.9}$ \\
	\hline
	& $0.6$  & $1$ & $0.676$ & $0.060$ & $11$ & $13$ & $14$ \\
	& & $5$ & $0.480$ & $0.112$ & $6$ & $7$ & $8$ \\
	&  & $10$ & $0.126$ & $0.140$ & $2$ & $3$ & $3$ \\
 	& $1$ & $1$ & $0.563$ & $0.088$ & $7$ & $8$ & $9$ \\
SmBP	&  & $5$ & $0.372$ & $0.141$ & $4$ & $5$ & $6$ \\
	&  & $10$ & $0.081$ & $0.144$ & $2$ & $2$ & $3$ \\
	& $1.4$ & $1$ & $0.463$ & $0.107$ & $5$ & $6$ & $7$ \\
	&  & $5$ & $0.299$ & $0.146$ & $4$ & $4$ & $5$ \\
	&  & $10$ & \textbf{0.088} & \textbf{0.174} & \textbf{2} & \textbf{2} & \textbf{2} \\
	\hline\hline
	& \multicolumn{2}{l}{\# clusters $G$} &  &  &  &  &  \\
	\hline
	KM & \multicolumn{2}{l}{$2$} & $0.377$ & $0.068$ & $-$ & $-$ & $-$ \\
	GM & \multicolumn{2}{l}{$2$} & $0.153$ & $0.106$ & $-$ & $-$ & $-$ \\
	& \multicolumn{2}{l}{$BIC$ selection} & $0.666$ & $0.034$ & $9$ & $10$ & $11$ \\
	\hline
\end{tabular}
\end{center}
\caption{Misclassification errors of SmBP clustering versus competitors and (when available) quantiles of the estimated number of clusters.}
\label{tab:bananas_misclas_err}
\end{table}

\subsection{Clustering: real data illustration}
\label{sec:clustering_real_data}

We illustrate how our clustering technique (from now on, SmBP{} clustering) works when applied to a real dataset. The aim is twofold: on one hand, it shows the cognitive support that the method could bring to the studied phenomenon and, on the other hand, what kind of practical problems could occur and how to treat them.

The presentation goes through three datasets belonging to different domains: spectrometric analysis, energy consumption and neuroscience.

\subsubsection{Spectrometric curves\label{sec:Tecator}}

Spectroscopic analysis is a fast, non--destructive and inexpensive technique which provides an estimate of the composition of an aliment based of the absorption of light emitted with different wavelengths by a spectrometer. Since the measure of absorption is a function of the wavelength, it represents a typical functional data.
In the last two decades, various functional techniques have been widely explored for this kind of data: see for instance, \cite{fer:vie06, del:hal:bat12} in the supervised classification framework.

\begin{figure}[tbp]
\begin{center}
	\includegraphics[height=0.28\textheight,width=0.32\textwidth]{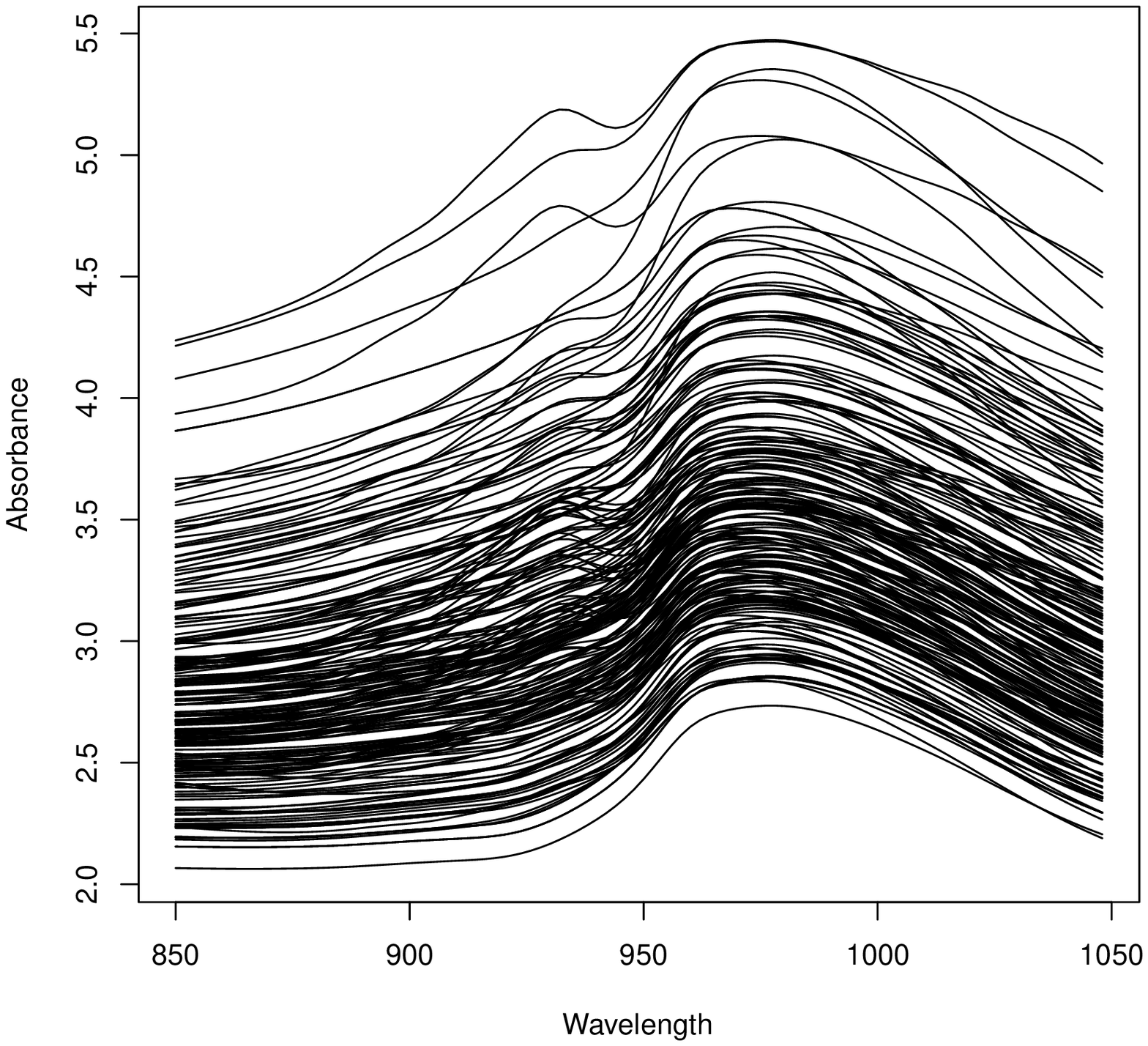} %
	\includegraphics[height=0.28\textheight,width=0.32\textwidth]{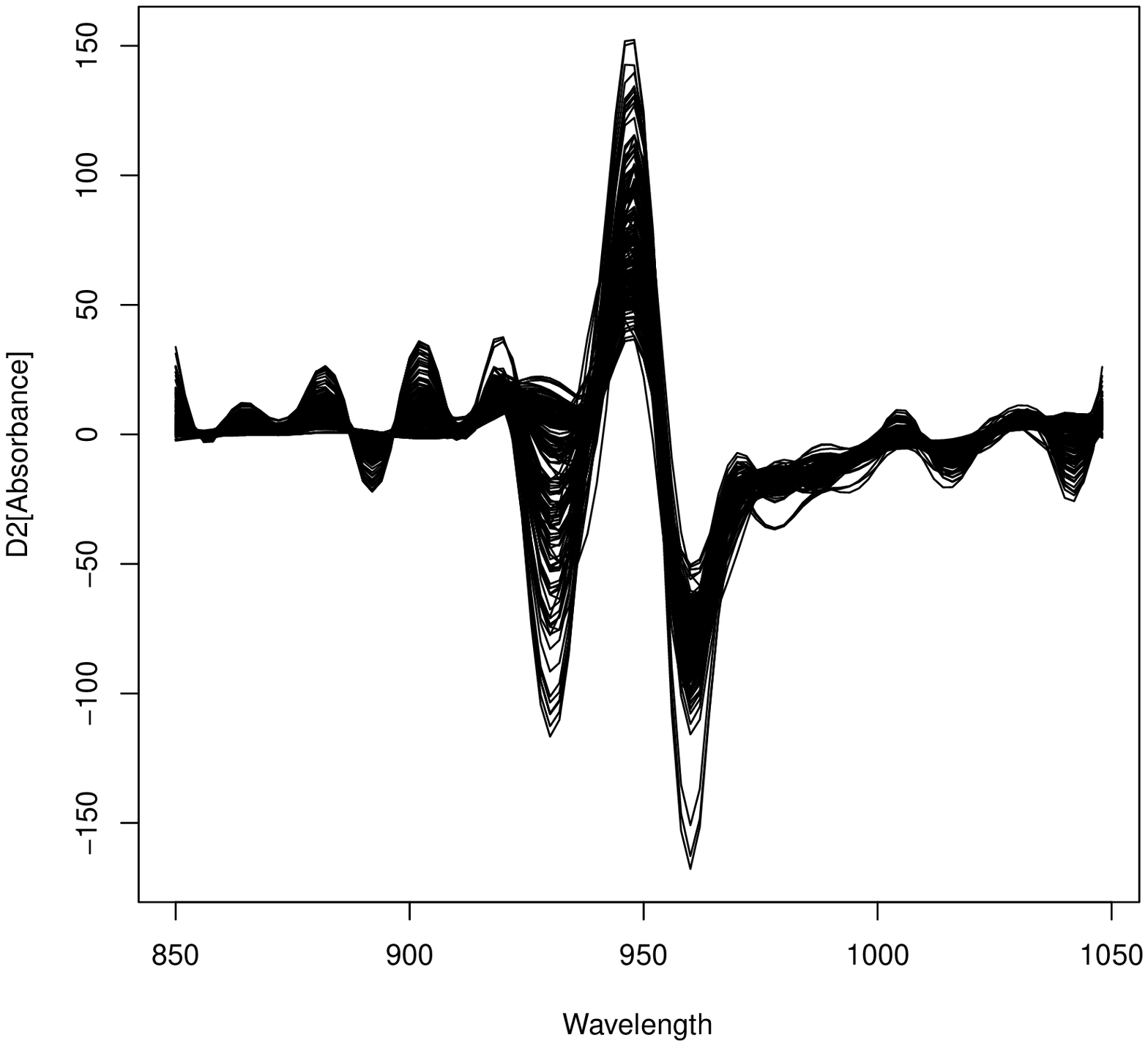} %
	\caption{Tecator curves (left) and their $2$-nd derivatives (right).}
	\label{fig:TecatorCurves}
\end{center}
\end{figure}

In the following, we illustrate an application of the SmBP{} clustering method to the well--known Tecator dataset (available at \url{http://lib.stat.cmu.edu/datasets/tecator}). It consists of 215 spectra in the near infra--red (NIR) wavelength range from 852 to 1,050 nm, discretized on a grid of 100 equispaced points, corresponding to the same number of finely chopped pork samples. Fat, protein, and water content, obtained by a traditional chemical analysis, is available for each sample. As conventionally done, in our study we consider the second derivatives of spectrometric curves instead of the original ones, to avoid the well-known \textquotedblleft calibration problem\textquotedblright due to the presence of shifts in the curves (see \citealp{fer:vie06}). Original spectrometric data and their second derivatives are visualized in Figure~\ref{fig:TecatorCurves}.

\begin{figure}[tbp]
	\begin{center}
	\includegraphics[height=0.28\textheight,width=0.32\textwidth]{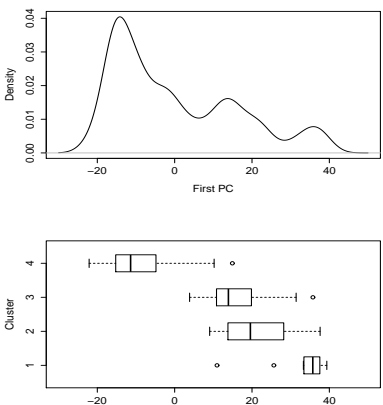}
	\caption{ Density estimate of the first PC of chemical components and the same stratified according to clusters analysis carried out with $r=5$ and $\protect\delta =0.2$.}
	\label{fig:TecatorClustering}
	\end{center}
\end{figure}

Since spectrometric data should represent a way to determine the chemical composition of the meat, the structure of the distribution of chemical components should be reproduced by the one of spectrometric curves. In this view, we first study the available chemical measures. The correlation analysis shows that the three components are highly linearly correlated: in particular, fat and water present a linear correlation coefficient equal to $-0.988$, whereas the content of protein exhibits a positive correlation with fat ($0.814$) and a negative one with water ($-0.861$). This suggests using PCA in order to summarize the chemical composition: in that way, the first PC explains the $98.5\%$ of the total variability. Observing the kernel estimate of the density of this PC (see the upper panel in Figure~\ref{fig:TecatorClustering}), the poly--modal distribution suggests that the sample is a mixture of three kinds of meats: the three groups are detected by considering the largest upper level sets containing the modes of the estimated density that, in the one dimensional case, reduces to look for the local minima whose abscissa identify class boundaries. 
Hence, it is expected that the spectrometric curve distribution presents a three modal structure as well as, that should be detected by the clustering algorithm defined in Section~\ref{sec:clustering_unsupervised_classification}.

After running a functional PCA on the second derivatives of spectrometric curves, we find that the spectrum is rather concentrated: the first three PCs explain $98.5\%$ of the total variability, and this suggests to use $d=3$ in our approximation. The selection of parameters $r$ and $\delta $ is performed according to the maximization of $CH$ index over a bivariate grid built with $r=1,\dots ,7$ and $\delta $ varying from $0.2$ to $1$ with step $0.1$. Index $CH$ reaches its maximum for $r=5$ and $\delta =0.2$, to which it corresponds $k=4$ clusters. In order to understand the appropriateness of this choice, besides the internal criterion  $CH$\ we use an external criterion by computing, for each couple of parameters $r,\delta $, the index of purity according to the three--group structure shown by the first PC which summarizes the chemical variables. It emerges that the couple $(r,\delta)$ maximizing $CH$ provides also a high degree of purity: this fact can be appreciated by inspecting Figure~\ref{fig:TecatorClustering}, and it provides a heuristic support of the possibility of reproducing the main features of the distribution of the chemical measures from the one of the spectrometric curves.

\subsubsection{District heating load-curves}

A district--heating system (or \textquotedblleft teleheating\textquotedblright ) allows the distribution of heat, generated in a centralized location, for entire districts through a network of insulated pipes. Due to its efficiency and to the pollution control, this system is spreading to many cities. In order to guarantee an optimal scheduling for generating heat, which allows choosing the right mix of on--line capacity, the analysis of the flows of heating demand is crucial. These flows depend mainly on two factors: an intra--daily pattern of the load demand, known as the load curve, and seasonal aspects.

To manage data from a district--heating system, also in a forecasting perspective, it is useful to stratify the set of load curves into a few homogeneous groups exhibiting similar demand patterns, since consumers characteristics are very different according to seasons and weather conditions. In what follows, we propose an application of our clustering algorithm to data on heat consumption in Turin, a northern Italian city, where the district heating is produced through a co--generation system.
The dataset has been used previously in a forecasting context based on regression approaches (see \citealp{goi:may:fus10,goi12}).

\begin{figure}[tbp]
	\begin{center}
	\includegraphics[height=0.28\textheight,width=0.32\textwidth]{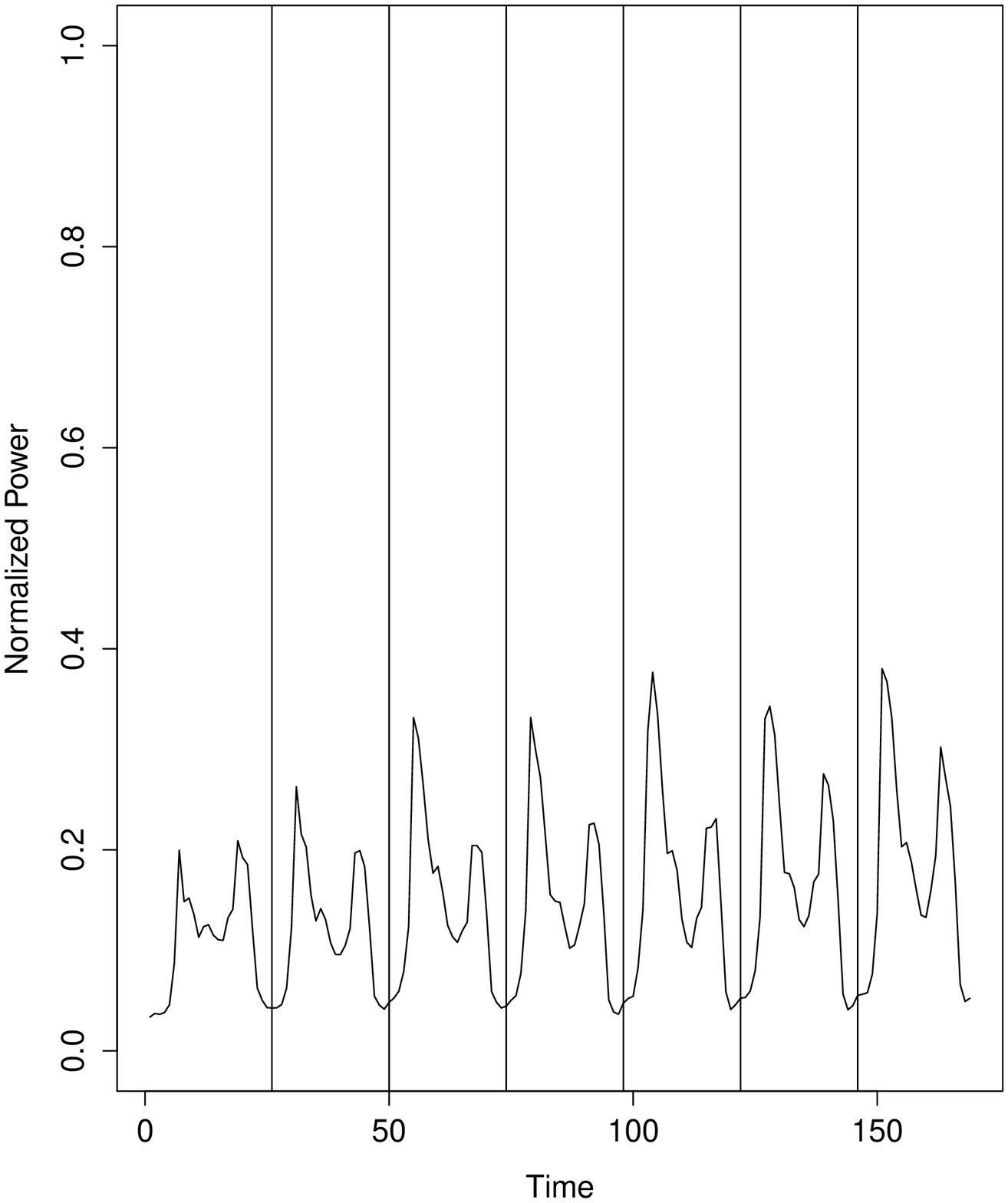} %
	\includegraphics[height=0.28\textheight,width=0.32\textwidth]{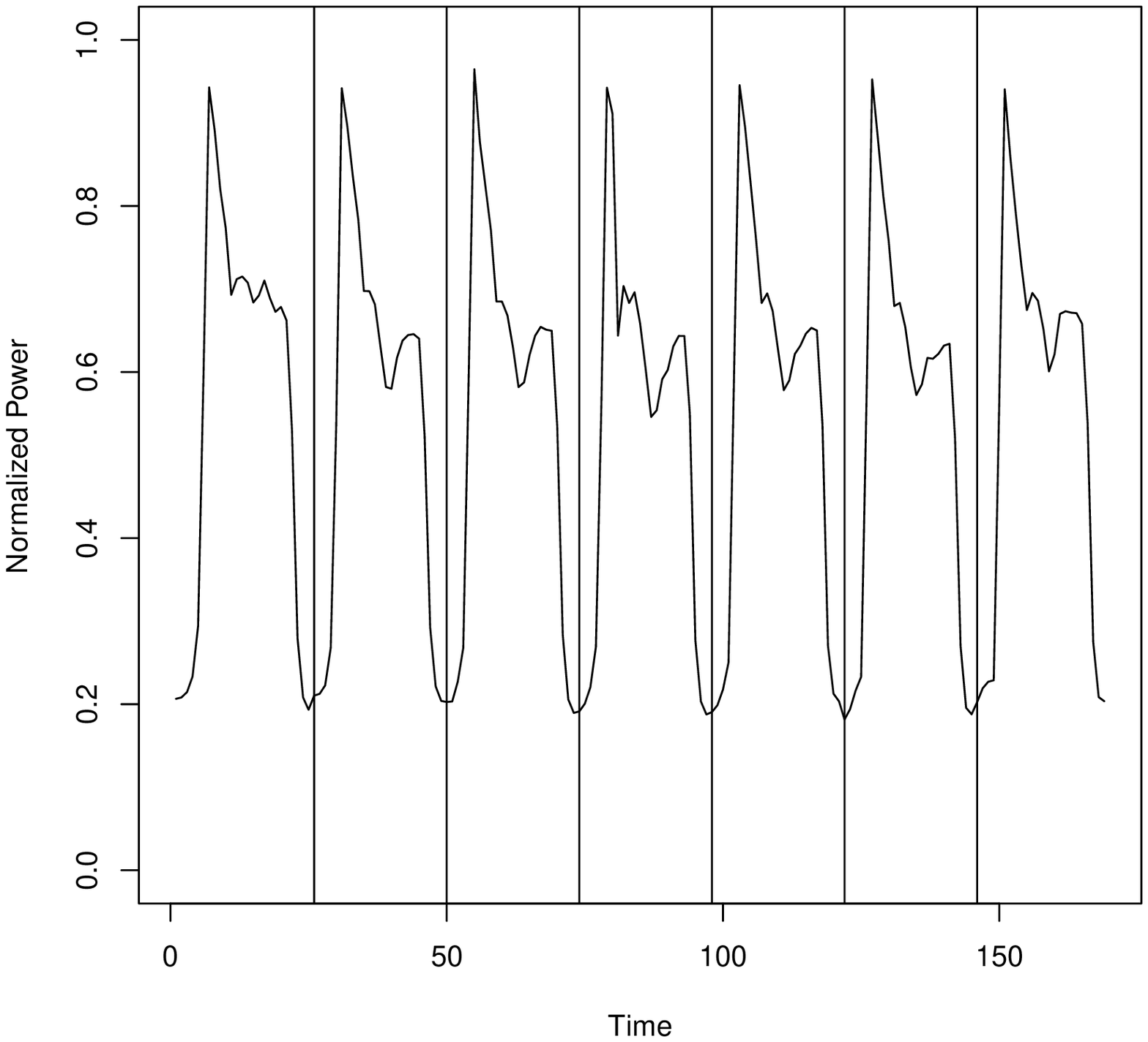} %
	\includegraphics[height=0.28\textheight,width=0.32\textwidth]{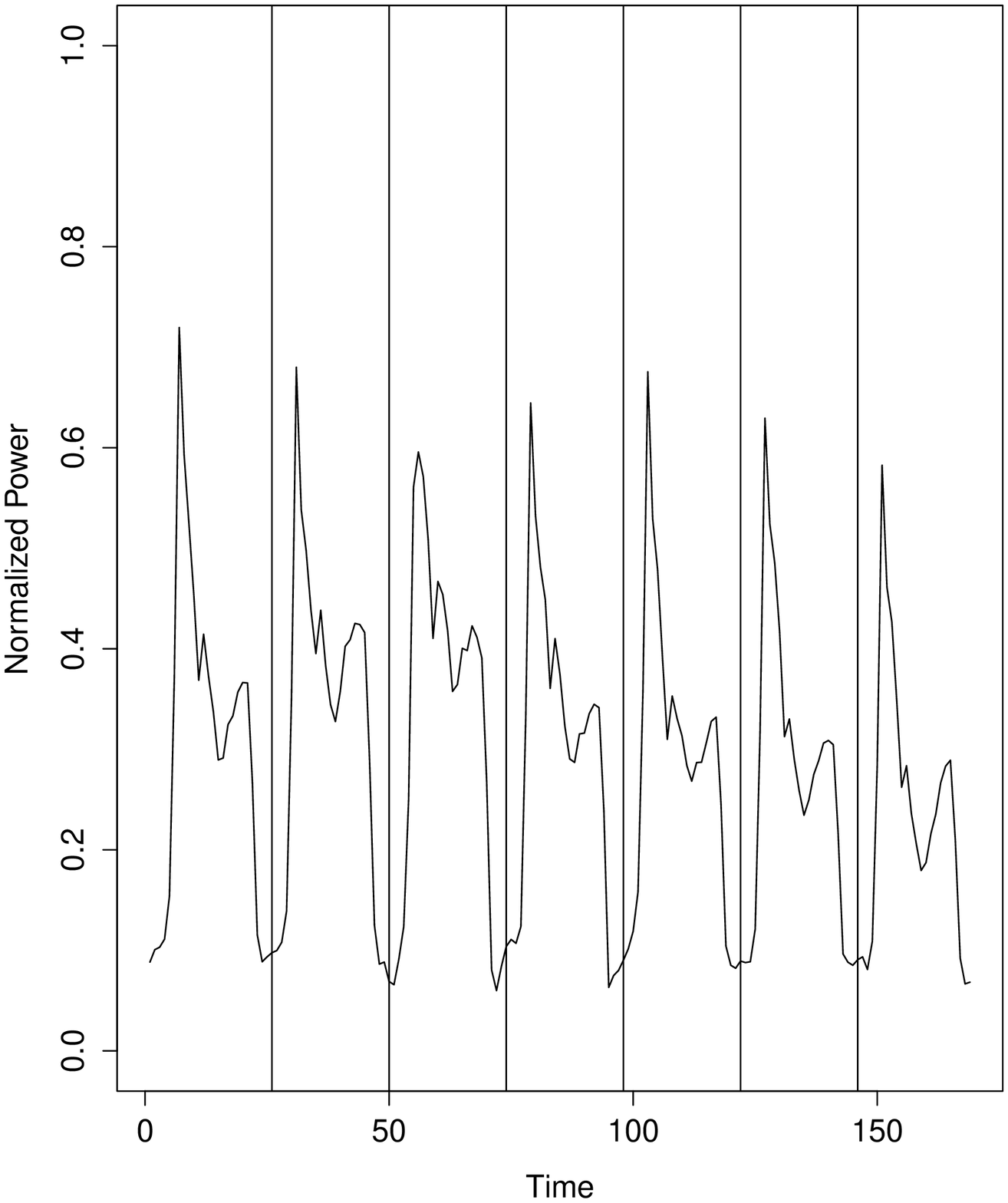}
	\caption{Demand of heat in a selected week of November 2002, January and
		March 2003.}
	\label{fig:LoadCurvesWeek}
	\end{center}	
\end{figure}

The dataset consists of hourly measurements of heat consumption for residential and commercial buildings during the periods October 15 -- April 20, covering the years 2001--02, 2002--03, 2003--04 and 2004--05. Due to privacy requests from the data supplier, the data have been normalized. Figure~\ref{fig:LoadCurvesWeek} displays the behaviour of the heating demand in three selected weeks in autumn, winter and spring: it is possible to distinguish the intra--daily pattern, due to an inertia in the demand reflecting the aggregate behaviour of consumers, as well as the seasonal evolution. Differently from electricity power demand, intra weekly differences among working days and weekends do not appear.

\begin{figure}[tbp]
	\begin{center}
	\includegraphics[height=0.28\textheight,width=0.32\textwidth]{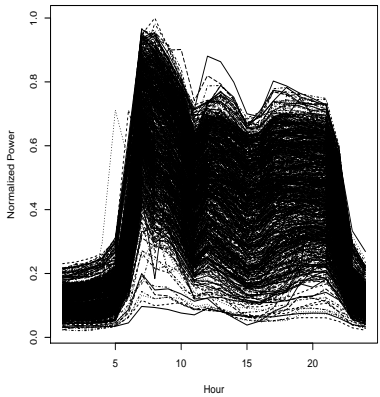}
	\caption{Normalized load curves, corresponding to daily profile.}
	\label{fig:LoadCurvesTot}
	\end{center}
\end{figure}

Taking advantage of the functional nature of the dataset, we split the series for each period into $187$ functional observations, each one coincident with a specific daily load curve. Finally, we dispose of a functional dataset consisting of $748$ curves discretized over an equispaced mesh of $24$ points. Figure~\ref{fig:LoadCurvesTot} displays the set of curves.

\begin{figure}[tbp]
	\includegraphics[height=0.28\textheight,width=0.32\textwidth]{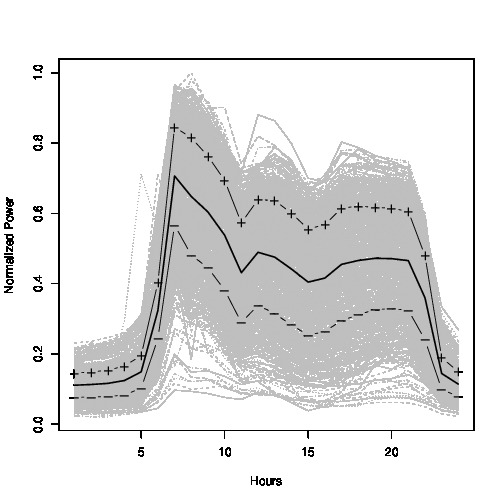} %
	\includegraphics[height=0.28\textheight,width=0.32\textwidth]{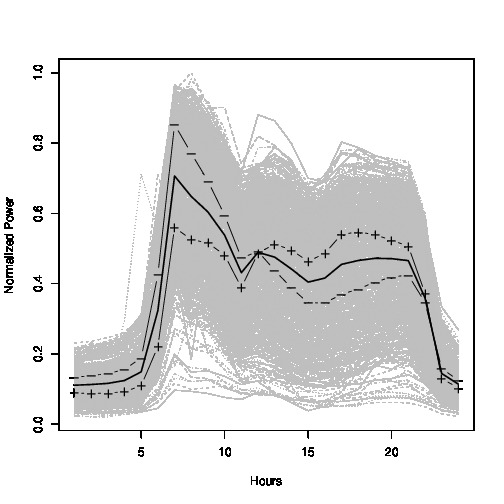} %
	\includegraphics[height=0.28\textheight,width=0.32\textwidth]{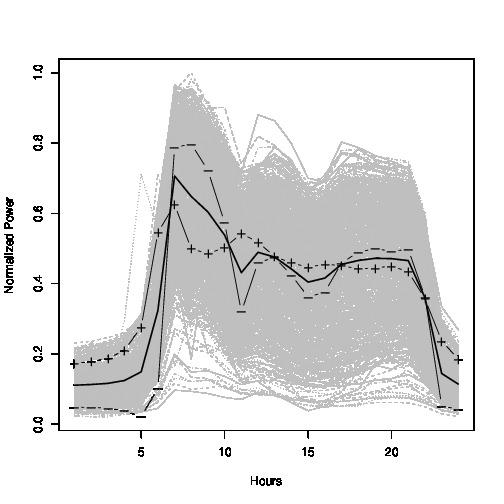}
	\caption{Contribution of the first three principal components.}
	\label{fig:LoadCurvesPCcontrib}
\end{figure}

The first step of our procedure is to perform a functional principal components analysis. It emerges that the spectrum is rather concentrated: the first three PCs explain more than the $97\%$ of the total variance, and thus it is sufficient to limit our analysis to $d\leq 3$. In order to provide an
interpretation of the contribution of the relevant PCs, we exploit a
graphical tool where we report the estimated mean curve plus and minus a
suitable multiple $M_{j}$ of each estimated eigenfunction: $\widehat{\mu }%
\pm M_{j}~\widehat{\xi }_{j}\sqrt{\widehat{\lambda }_{j}}$ (see e.g.~\citealp{ram:sil05}). The results, visualized in Figure~\ref{fig:LoadCurvesPCcontrib}, show that the first eigenfunction, which does not present sign changes, describes a vertical shift effect, due to weather
conditions in seasons, whereas the second eigenfunction highlights
differences among demand in the morning and in the remaining part of the
day: it seems to be related to the heat retention ability of buildings (a greater
heating in the morning produces less need in the afternoon). Finally, the
third eigenfunction seems to be connected and counter--posed to the three peaks of
demand that appear systematically during the day in the morning, in the
afternoon and in the evening.

\begin{figure}[tbp]
\begin{center}
	\includegraphics[height=0.28\textheight,width=0.32\textwidth]{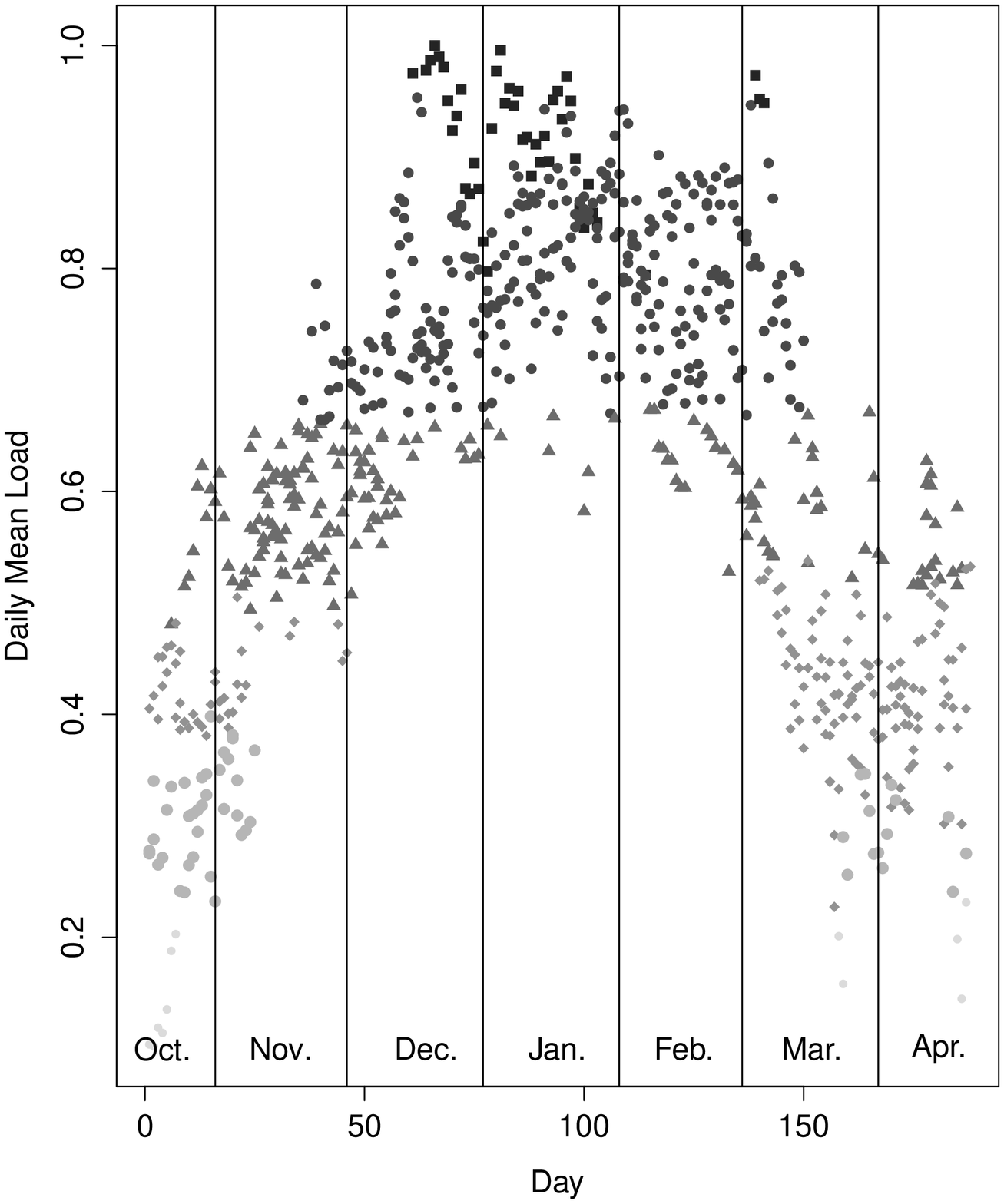} %
	\includegraphics[height=0.28\textheight,width=0.32\textwidth]{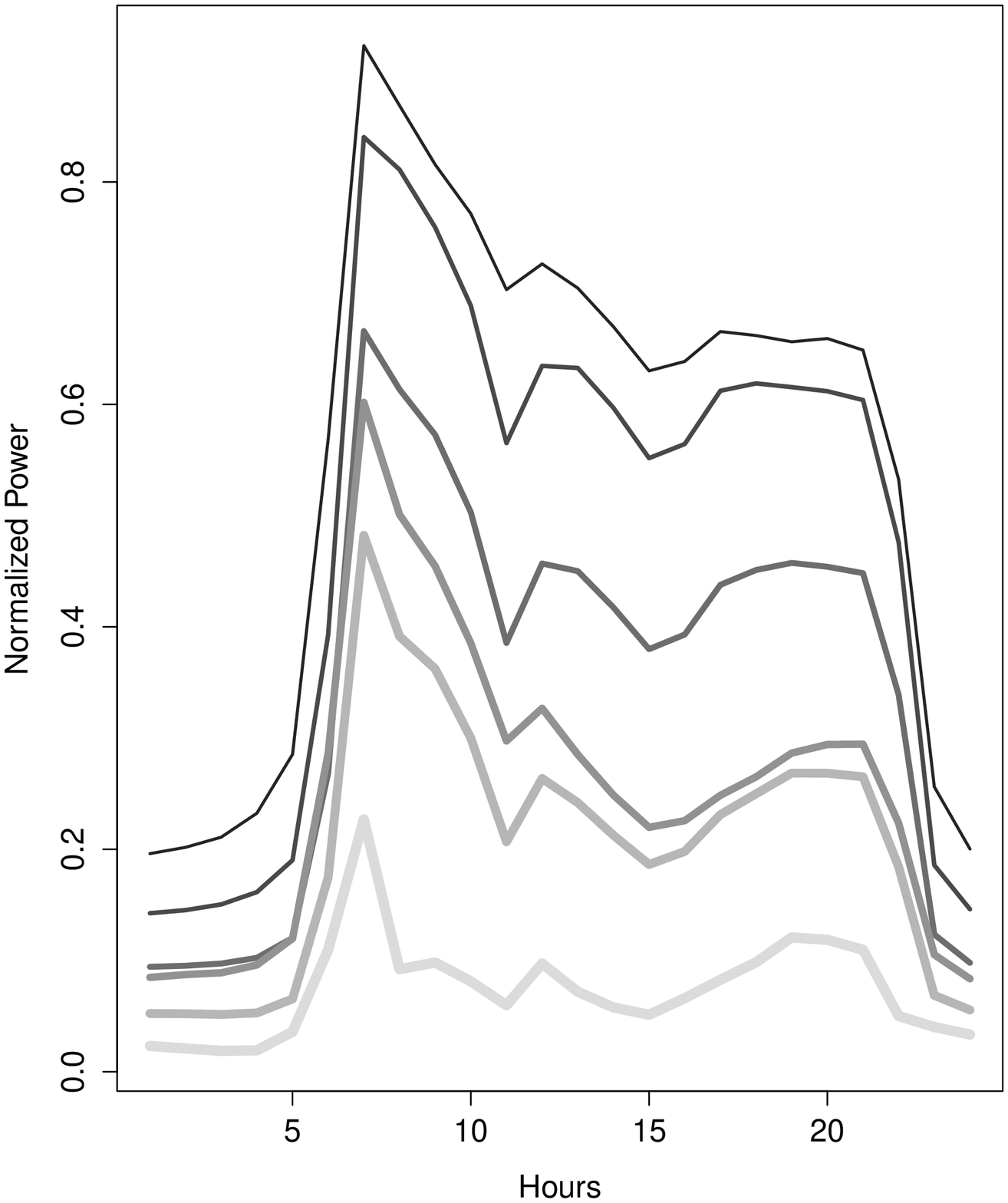} %
	\includegraphics[height=0.28\textheight,width=0.32\textwidth]{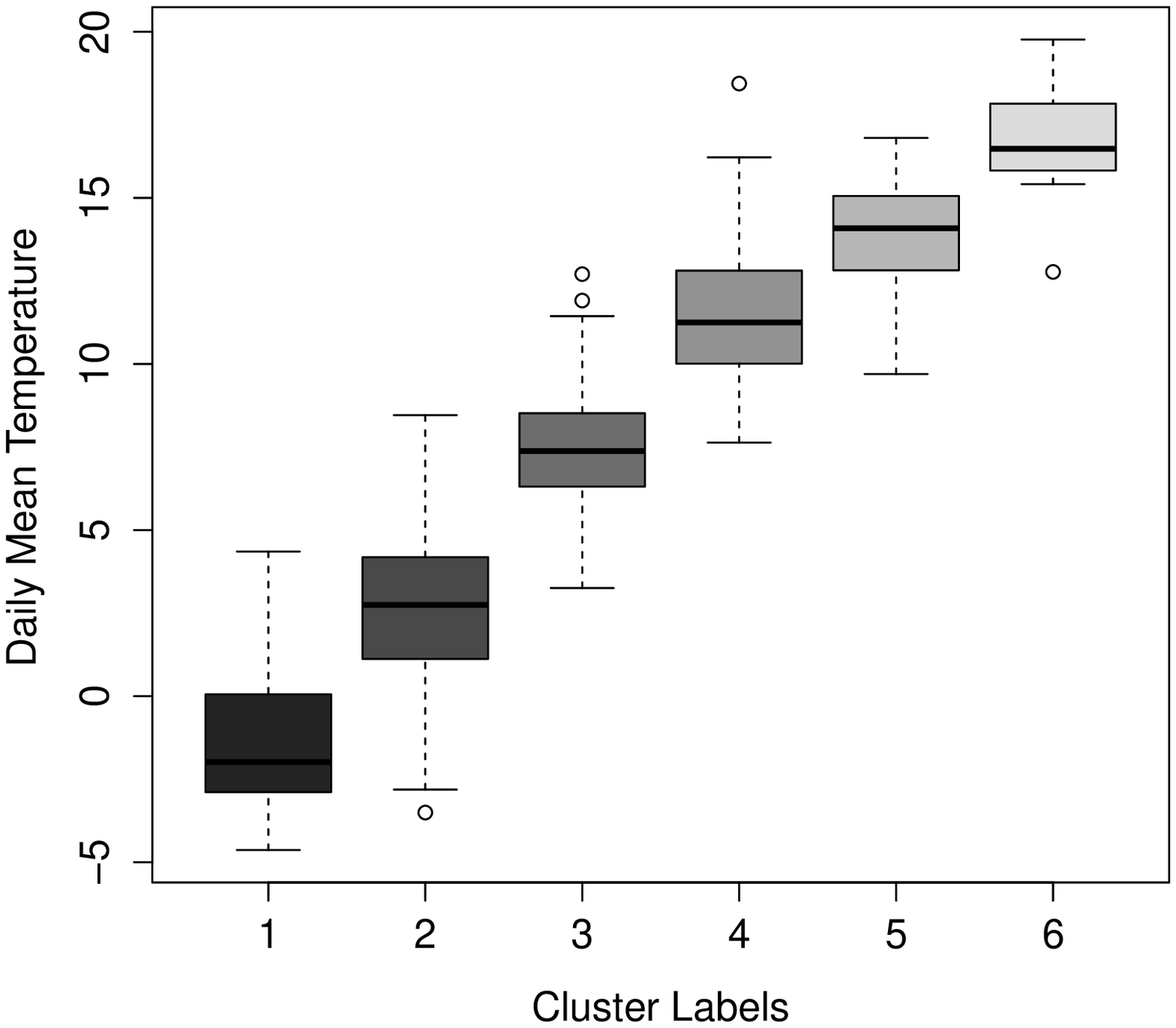}
	\caption{Calendar positioning of clusters against the daily mean load, corresponding modal curves and relationship among clusters and daily mean temperature. Along the panels, a cluster is identified by the same grey level.}
	\label{fig:LoadCurvesClustering}
\end{center}
\end{figure}

In order to apply the SmbP{} cluster algorithm, one has to preventively select the parameters $r$ and $\delta $. Using the same grid as in Section \ref{sec:Tecator}, the $CH$ criterion suggests $r=3$ and $\delta =0.5$, a choice that leads to $k=6$ clusters. These clusters reflect the differences in level and behaviour of daily demand of heating in the different seasons: high levels of demand in winter with a strong peak in the morning, moderate level in autumn and spring with load curves presenting three peaks (in the morning, in the afternoon and in the evening). To better understand the effect of clustering, in Figure~\ref{fig:LoadCurvesClustering} we report the calendar positioning of each element of the clusters (each point represents a specific load curves, synthesized by its daily average) and alongside the modal curves, plotted using the same level of grey. We also report the box-plots resulting after a stratification of the daily mean temperature by the cluster labels: one can note that the temperature, without presenting a multi--modal density, is one of the most important external variables that can be used in such a clustering exercise.
Matching the results, we recognize the typical patterns for winter and mid-seasons, distinguish freezing, cold and mild days. Finally, it emerges that, if one was to set up a forecasting model, an accurate prediction of temperature would be the key to making accurate prediction in the demand of heating.

\subsubsection{Neuronal experiment} 

The analysis of neuronal spiking activity, recorded under various behavioural
conditions, is a central tool in neuroscience: data acquired from multiple
neurons are essential to explain neural information processing. The problem
is that contributions of multiple cells must be disentangled from the
background noise and from each other in order to analyze the activity of
individual neurons. The procedure that allows distinguishing the activity of
one or more neurons from a noisy time series is known as spike sorting.

\begin{figure}[tbp]
	\begin{center}
	\includegraphics[height=0.28\textheight,width=0.32\textwidth]{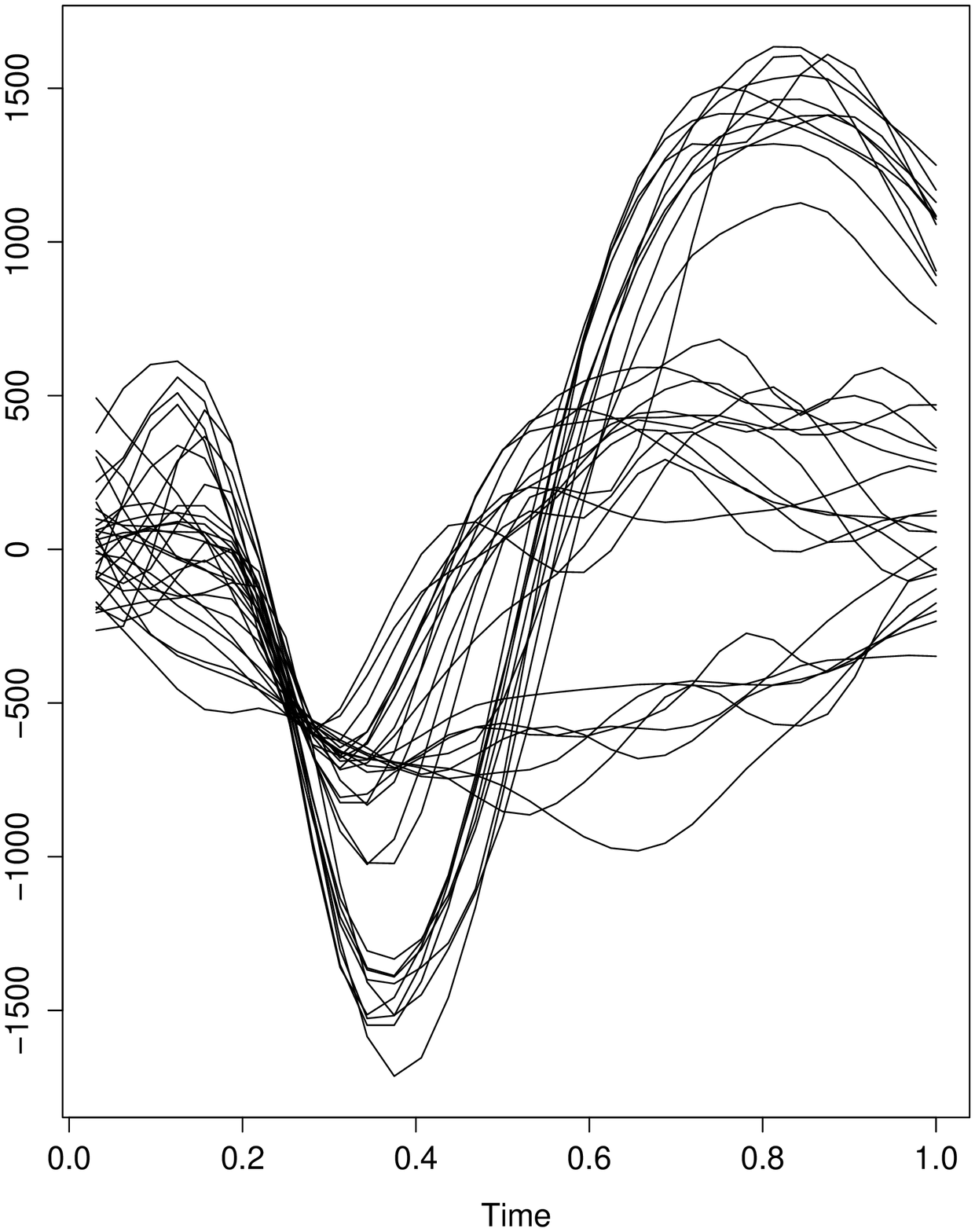}
	\includegraphics[clip=true, trim = 7cm 2cm 7cm 2cm, height=0.28\textheight,width=0.32\textwidth]{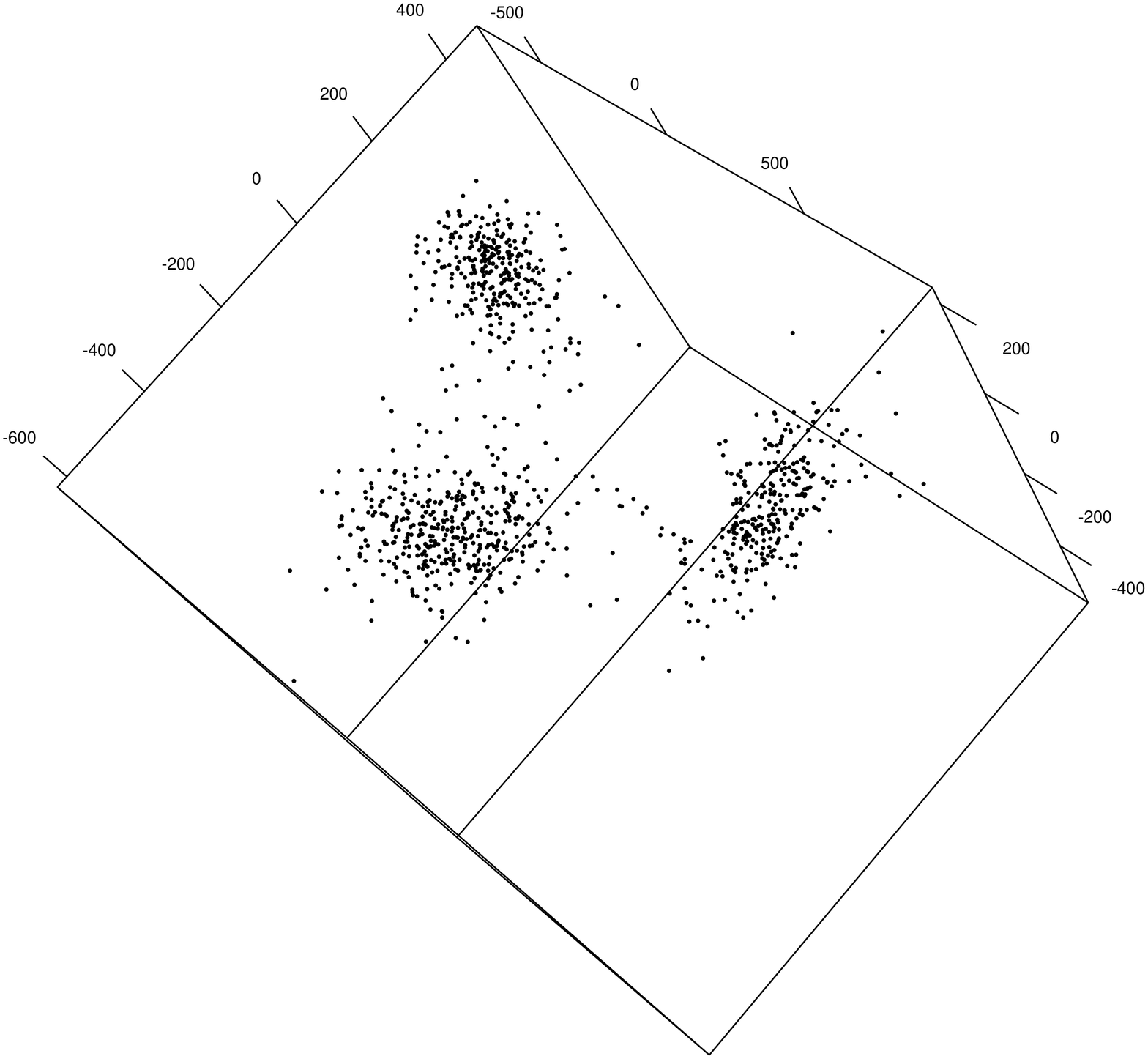} 
	\includegraphics[clip=true, trim = 7cm 2cm 7cm 2cm, height=0.28\textheight,width=0.32\textwidth]{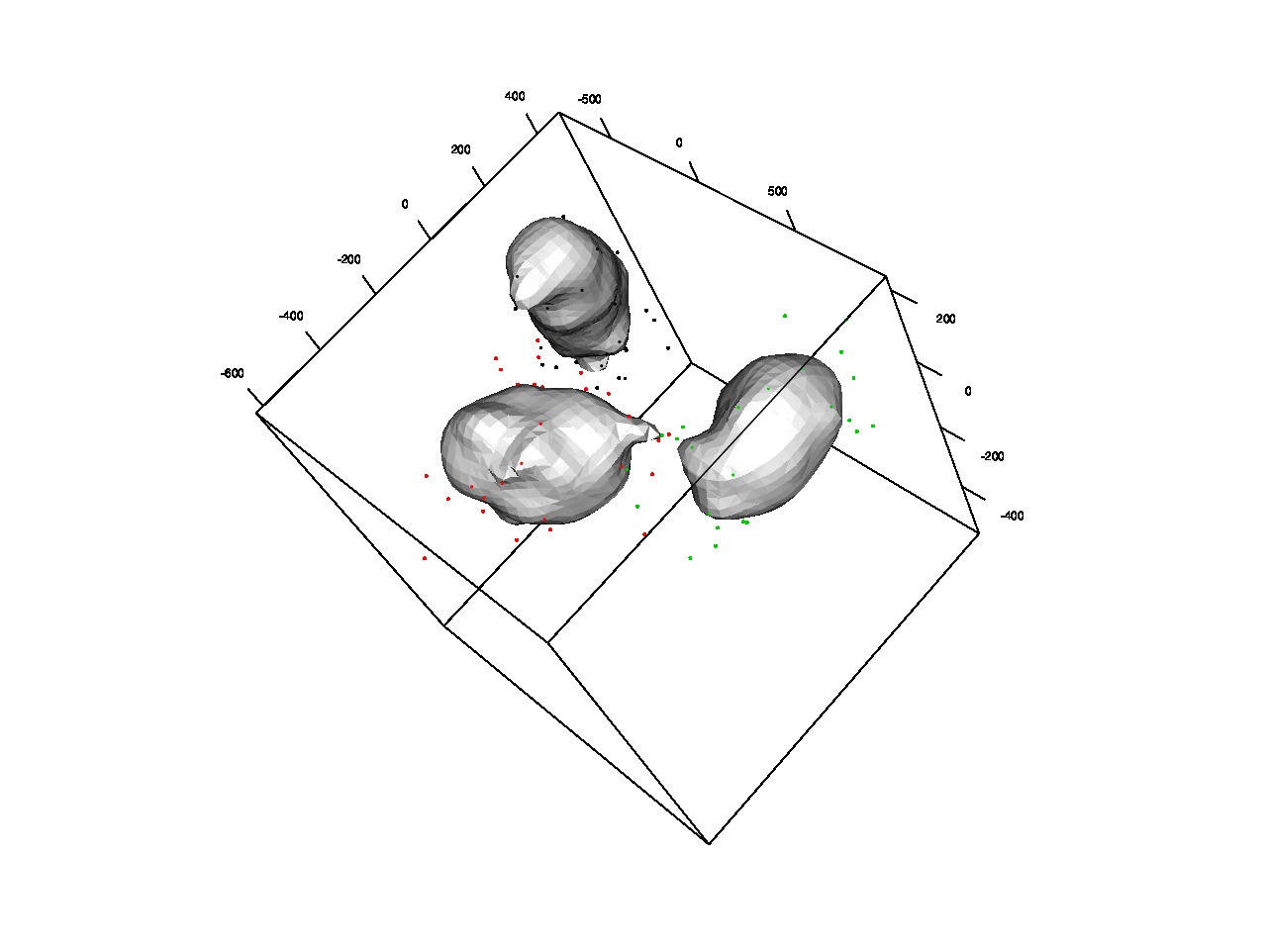}
	\caption{Neuronal Experiment (left to right): a random sample of 30 curves, 3-D Scatter plot of the first three Principal Components and the corresponding maximal level set when $r=7$ and $\protect\delta =0.8$.}
	\label{fig:SpikesCloud}
	\end{center}
\end{figure}

In this section we show how the SmBP clustering can contribute to the spike sorting: each detected cluster can be thought to correspond to the activity of a single neuron. The dataset comes from a behavioural experiment performed at the Andrew Schwartz motorlab (\url{http://motorlab.neurobio.pitt.edu/index.php}) on a macaque monkey performing a center-out and out-center target reaching task with $26$ targets in a virtual $3D$ environment (see \citealp{tod:sad:bat:cha:ven14} for a detailed description of the experiment considered). The neural activity recorded consists of all the action potentials detected above a channel-specific threshold on a $96$-channel Utah array implanted in the primary motor cortex. The data set is split into $1000$ functional data representing the voltage of neurons versus the time, discretized over a grid of $32$ equispaced time points (normalized between $0$ and $1$). A sample, of $30$ selected randomly curves, is shown in the left panel of Figure~\ref{fig:SpikesCloud}. An analysis of (a larger set of) these curves can also be found in \cite{tod:sad:bat:cha:ven14}.

Performing the functional PCA on such a set of curves, we observe that also in this case the spectrum is concentrated: the first PC explains $78.6\%$ of the total variability, and the explained variance by the first three PCs is about $96.4\%$. Observing the 3-dimensional scatterplot of the first three PCs (see Figure~\ref{fig:SpikesCloud}, left panel), three clouds appear evidently.

In order to detect a good choice for the parameters $r$ and $\delta $, we perform our clustering algorithm with $d=3$ over a grid built from $r=2,\dots ,7$ (with step equal to $1$), and $\delta =0.2,\dots ,1$ (with step $0.1$) and compute the $CH$ indexes. The analysis of the obtained values leads to various admissible configurations for the couple $\left( r,\delta \right) $, to which there always correspond $k=3$ clusters: for instance $r=3,\dots ,7$ combined with $\delta =0.8,0.9,1$ produce the same number of clusters and the maximal $CH$. The middle panel of Figure~\ref{fig:SpikesCloud} shows the maximal level set when one uses $r=7$ and $\delta =0.8$.

\begin{figure}[tbp]
\begin{center}
	\includegraphics[height=0.28\textheight,width=0.32\textwidth]{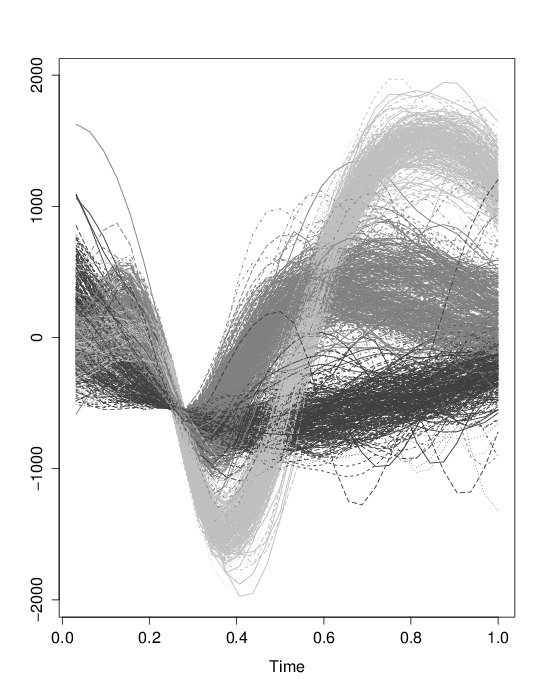} %
	\includegraphics[height=0.28\textheight,width=0.32\textwidth]{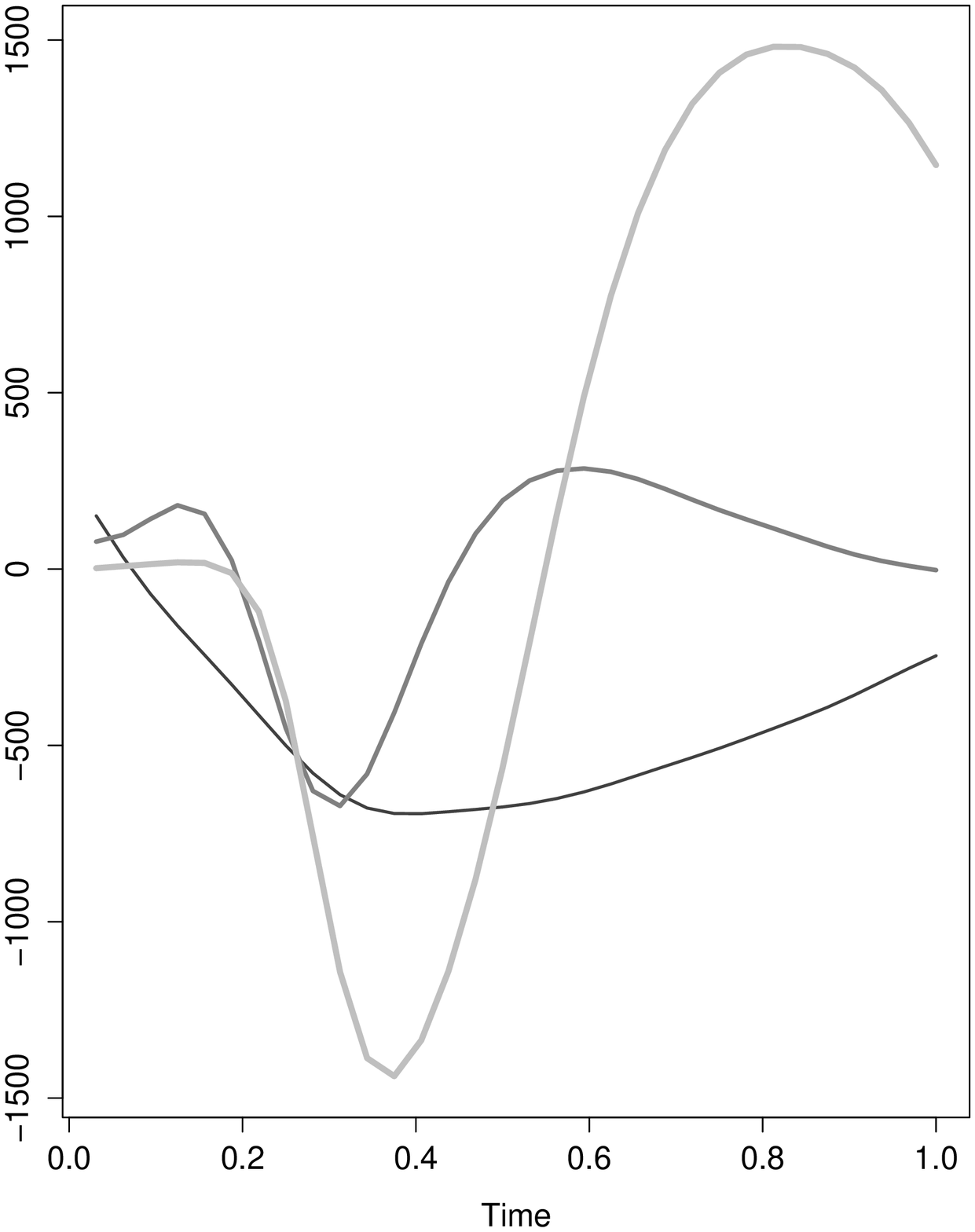}
	\caption{Neuronal Experiment: clusters and modal curves when $r=7$ and $\protect\delta =0.8$.}
	\label{fig:SpikesCluster}	
\end{center}
\end{figure}

The result of the clustering procedure is visualized in Figure~\ref{fig:SpikesCluster},
where the centers of prototypes (the modal curves) and the corresponding
clusters are reproduced.

\subsection{Discriminant: simulation and real data illustration}
\label{sec:discriminant_simulation_real_data}

The aim of this section is to assess the performance of the supervised classification algorithm illustrated in Section~\ref{sec:discriminant_supervised_classification} (briefly, SmBP{} classifier) by the analysis of both simulated and real datasets. The experiments consist in computing the (empirical) distribution out--of--sample misclassification error by a two--fold cross--validation procedure repeated $100$ times: more in details, for each available sample, at each iteration, $2/3$ of the data are used in evaluating the classifier, and the misclassification error is estimated on the remaining part. The estimation of the density in each group is performed by a multivariate kernel density estimator with $H$ diagonal, selected according to Section~\ref{sec:surrogate_density_estimation_and_H_choice}.

According to the comments in Section~\ref{sec:parameters_tuning}, classifiers are implemented with $d=2,\ldots,5$, where $d=5$ is considered only to see how the results worsen due to the curse of dimensionality. The out--of--sample errors are compared with those computed by using parametric and nonparametric competitors: the GLM classifier based on the coefficients of a basis representation for functional data, nonparametric discrimination using kernel (NP in the following) and k--NN estimation, based on the classic $L^{2} $ metric (see \citealp{fer:vie06}). All computations are done with the software R: in particular, the competitor algorithms are taken from the package \textit{fda.usc} (see \citealp{feb:ovi12}).

\subsubsection{Analysis of simulated data}

The dataset are generated according to the two ``horseshoes'' model described in Section~\ref{sec:clustering_simulations}, with the following settings:
\begin{enumerate}[-]
	
	\item sample size $n=150, 300, 450$ with training--sets of size $n_{in}=100,200,300$;
	
	\item we consider both the balanced case ($\pi_1=1/2$), and two unbalanced cases (with $\pi_1=1/3$ and $\pi_1=1/4$);
	
	\item the vertical translation parameter $k$ equals $0.5$;
	
	\item three degrees of variability are considered to model the noise around the two semi--circumferences: $\sigma = 0.05, 0.10, 0.15,$ (small, medium and high variability).
	
\end{enumerate}
\begin{table}[tbp]
	\begin{center}
		\begin{tabular} {cccccccccc}
			\hline
			$n_{in}$ & $\sigma $ & $n_{out}$ &  $d=2$  &  $d=3$  &  $d=4$  &  $d=5$  &   GLM   &  k--NN  &   NP    \\ \hline
			         &   0.05    &           &  0.155  &  0.026  &  0.039  &  0.058  &  0.361  &  0.024  &  0.042  \\
			         &           &           & (0.039) & (0.015) & (0.024) & (0.035) & (0.060) & (0.020) & (0.023) \\
			  100    &   0.10    &    50     &  0.228  &  0.095  &  0.111  &  0.134  &  0.345  &  0.104  &  0.130  \\
			         &           &           & (0.052) & (0.032) & (0.040) & (0.044) & (0.054) & (0.041) & (0.041) \\
			         &   0.15    &           &  0.257  &  0.188  &  0.195  &  0.221  &  0.397  &  0.159  &  0.200  \\
			         &           &           & (0.061) & (0.045) & (0.050) & (0.051) & (0.064) & (0.049) & (0.048) \\ \hline
			         &   0.05    &           &  0.185  &  0.033  &  0.042  &  0.057  &  0.307  &  0.041  &  0.047  \\
			         &           &           & (0.042) & (0.017) & (0.017) & (0.019) & (0.031) & (0.018) & (0.022) \\
			  200    &   0.10    &    100    &  0.275  &  0.150  &  0.154  &  0.167  &  0.335  &  0.136  &  0.148  \\
			         &           &           & (0.053) & (0.049) & (0.027) & (0.028) & (0.029) & (0.038) & (0.031) \\
			         &   0.15    &           &  0.255  &  0.228  &  0.234  &  0.265  &  0.440  &  0.232  &  0.250  \\
			         &           &           & (0.039) & (0.033) & (0.031) & (0.035) & (0.047) & (0.040) & (0.036) \\ \hline
			         &   0.05    &           &  0.157  &  0.020  &  0.033  &  0.036  &  0.356  &  0.030  &  0.035  \\
			         &           &           & (0.029) & (0.007) & (0.011) & (0.010) & (0.033) & (0.010) & (0.011) \\
			  300    &   0.10    &    150    &  0.217  &  0.100  &  0.117  &  0.128  &  0.383  &  0.093  &  0.108  \\
			         &           &           & (0.033) & (0.024) & (0.023) & (0.026) & (0.036) & (0.026) & (0.026) \\
			         &   0.15    &           &  0.234  &  0.171  &  0.183  &  0.192  &  0.411  &  0.162  &  0.178  \\
			         &           &           & (0.030) & (0.024) & (0.026) & (0.023) & (0.030) & (0.026) & (0.027) \\ \hline
		\end{tabular}		
	\end{center}
	\caption{Estimated mean and standard deviation (in parentheses) of
		misclassification error for the two--horseshoes setting. Two balanced groups: $\pi_1=1/2$.}
	\label{tab:MisscalssErrBananas1}
\end{table}
\begin{table}[tbp]
	\begin{center}
		\begin{tabular}{cccccccccc} 
			\hline
			$n_{in}$ & $\sigma $ & $n_{out}$ &  $d=2$  &  $d=3$  &  $d=4$  &  $d=5$  &   GLM   &  k--NN  &   NP    \\ \hline
			         &   0.05    &           &  0.156  &  0.051  &  0.067  &  0.069  &  0.342  &  0.036  &  0.060  \\
			         &           &           & (0.045) & (0.029) & (0.038) & (0.039) & (0.067) & (0.023) & (0.039) \\
			  100    &   0.10    &    50     &  0.208  &  0.154  &  0.154  &  0.172  &  0.287  &  0.101  &  0.163  \\
			         &           &           & (0.044) & (0.048) & (0.047) & (0.045) & (0.057) & (0.043) & (0.043) \\
			         &   0.15    &           &  0.241  &  0.212  &  0.225  &  0.252  &  0.366  &  0.175  &  0.221  \\
			         &           &           & (0.050) & (0.048) & (0.052) & (0.061) & (0.058) & (0.048) & (0.050) \\ \hline
			         &   0.05    &           &  0.150  &  0.035  &  0.034  &  0.045  &  0.340  &  0.030  &  0.034  \\
			         &           &           & (0.045) & (0.013) & (0.014) & (0.015) & (0.036) & (0.014) & (0.015) \\
			  200    &   0.10    &    100    &  0.211  &  0.133  &  0.153  &  0.175  &  0.342  &  0.129  &  0.146  \\
			         &           &           & (0.038) & (0.028) & (0.029) & (0.036) & (0.038) & (0.030) & (0.029) \\
			         &   0.15    &           &  0.263  &  0.180  &  0.188  &  0.201  &  0.359  &  0.169  &  0.193  \\
			         &           &           & (0.038) & (0.035) & (0.033) & (0.032) & (0.042) & (0.036) & (0.034) \\ \hline
			         &   0.05    &           &  0.137  &  0.036  &  0.039  &  0.051  &  0.327  &  0.039  &  0.043  \\
			         &           &           & (0.025) & (0.010) & (0.012) & (0.015) & (0.042) & (0.015) & (0.013) \\
			  300    &   0.10    &    150    &  0.155  &  0.083  &  0.098  &  0.100  &  0.317  &  0.080  &  0.096  \\
			         &           &           & (0.028) & (0.020) & (0.021) & (0.022) & (0.031) & (0.021) & (0.021) \\
			         &   0.15    &           &  0.200  &  0.167  &  0.170  &  0.182  &  0.341  &  0.154  &  0.185  \\
			         &           &           & (0.031) & (0.029) & (0.028) & (0.028) & (0.038) & (0.028) & (0.031) \\ \hline
		\end{tabular}
	\end{center}
	\caption{Estimated mean and standard deviation (in parentheses) of
		misclassification error for the two--horseshoes setting. Two unbalanced groups: $\pi_1=1/3$.}
	\label{tab:MisscalssErrBananas2}	
\end{table}
\begin{table}[tbp]
	\begin{center}
		\begin{tabular}{cccccccccc} 
			\hline
			$n_{in}$ & $\sigma $ & $n_{out}$ &  $d=2$   &  $d=3$  &  $d=4$  &  $d=5$  &   GLM   &  k--NN  &   NP    \\ \hline
			         &   0.05 &           &  0.112   &  0.042  &  0.046  &  0.064  &  0.279  &  0.018  &  0.040  \\
			         &           &           & (0.048)  & (0.031) & (0.031) & (0.042) & (0.070) & (0.019) & (0.035) \\
			  100  &   0.10 &    50    &  0.183   &  0.151  &  0.156  &  0.164  &  0.266  &  0.082  &  0.145  \\
			         &           &           & (0.044 ) & (0.047) & (0.047) & (0.046) & (0.054) & (0.041) & (0.048) \\
			         &   0.15  &           &  0.169   &  0.148  &  0.142  &  0.161  &  0.260  &  0.092  &  0.147  \\
			         &           &           & (0.042)  & (0.042) & (0.047) & (0.049) & (0.056) & (0.047) & (0.040) \\ \hline
			         &   0.05 &           &  0.098   &  0.024  &  0.031  &  0.033  &  0.252  &  0.020  &  0.032  \\
			         &           &           & (0.021)  & (0.013) & (0.011) & (0.013) & (0.037) & (0.015) & (0.018) \\
			  200 &   0.10  &    100  &  0.152   &  0.121  &  0.128  &  0.138  &  0.258  &  0.084  &  0.120  \\
			         &           &           & (0.040)  & (0.033) & (0.033) & (0.032) & (0.035) & (0.025) & (0.032) \\
			         &   0.15  &           &  0.181   &  0.152  &  0.170  &  0.179  &  0.271  &  0.129  &  0.150  \\
			         &           &           & (0.034)  & (0.031) & (0.031) & (0.031) & (0.030) & (0.030) & (0.032) \\ \hline
			         &   0.05 &           &  0.119   &  0.013  &  0.021  &  0.033  &  0.271  &  0.018  &  0.020  \\
			         &           &           & (0.020)  & (0.009) & (0.013) & (0.016) & (0.033) & (0.008) & (0.008) \\
			  300 &   0.10  &    150 &  0.153   &  0.127  &  0.131  &  0.128  &  0.265  &  0.107  &  0.139  \\
			         &           &           & (0.028)  & (0.024) & (0.025) & (0.025) & (0.023) & (0.024) & (0.025) \\
			         &   0.15  &           &  0.174   &  0.169  &  0.181  &  0.194  &  0.256  &  0.156  &  0.181  \\
			         &           &           & (0.027)  & (0.028) & (0.025) & (0.027) & (0.028) & (0.024) & (0.025) \\ \hline
		\end{tabular}
	\end{center}
	\caption{Estimated mean and standard deviation (in parentheses) of
		misclassification error for the two--horseshoes setting. Two unbalanced groups: $\pi_1=1/4$.}
	\label{tab:MisscalssErrBananas3}	
\end{table}
\begin{figure}[tbp]
	\includegraphics[height=0.28\textheight,width=0.32\textwidth]{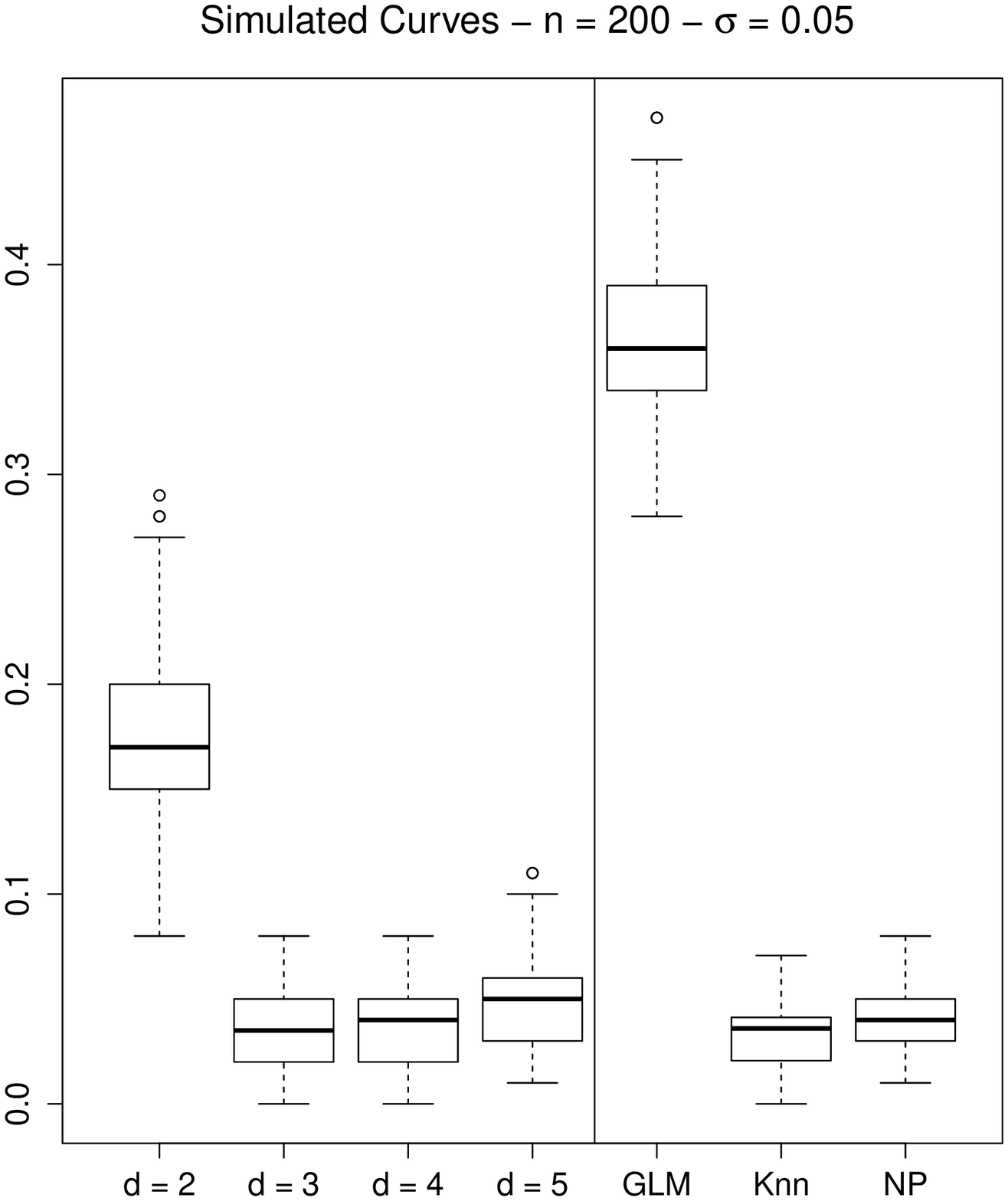}
	\includegraphics[height=0.28\textheight,width=0.32\textwidth]{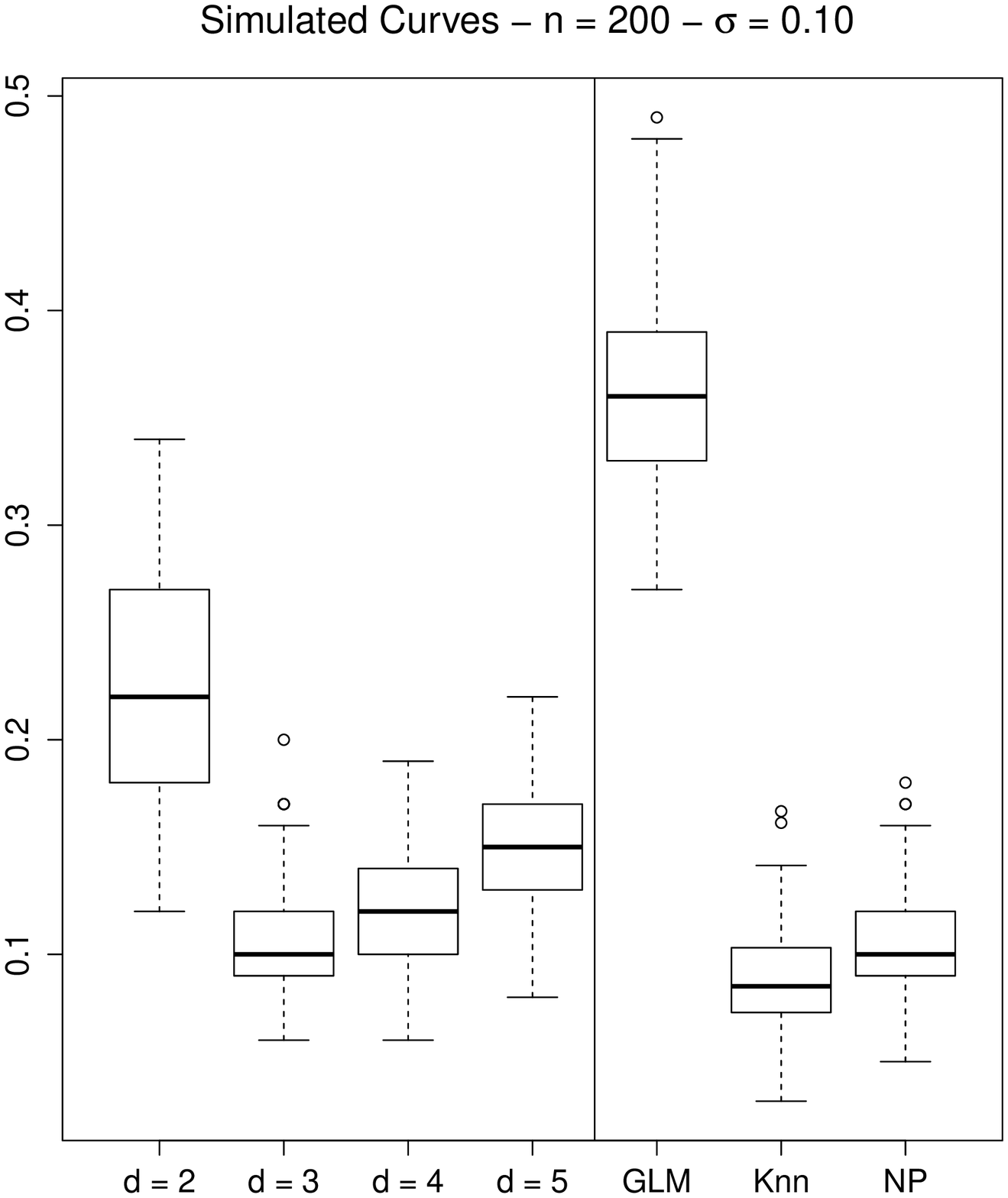}
	\includegraphics[height=0.28\textheight,width=0.32\textwidth]{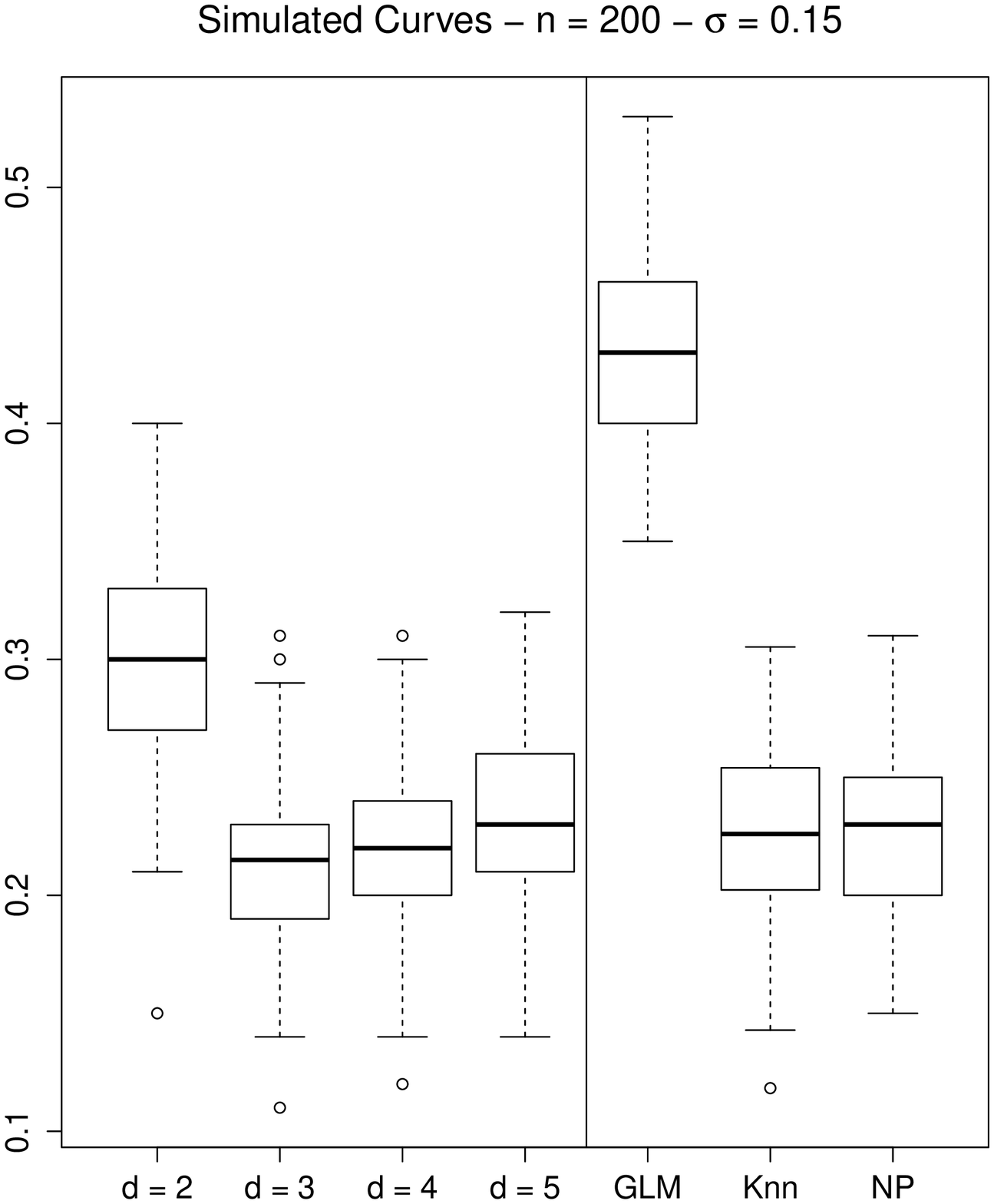}
	\caption{Distributions of misclassification errors estimated over $100$
		replications when $n_{in}=200$ and $\protect\sigma =0.05,0.10,0.15$ (form left to
		right).}
	\label{fig:BananeDiscriminant}
\end{figure}
After performing the discrimination exercise, one obtains the misclassification error distributions whose summary measures (mean and standard deviation) are collected in tables~\ref{tab:MisscalssErrBananas1}, \ref{tab:MisscalssErrBananas2} and \ref{tab:MisscalssErrBananas3}. In Figure~\ref{fig:BananeDiscriminant} such error distributions when $n_{in}=200$, $\sigma = 0.05 , 0.10, 0.15$ are reproduced. 
\\
From the tables and plots, it emerges that the SmBP classifier produces the best results, both in the balanced and unbalanced cases when $d=3$. This is coherent with the fact that $FEV(3)\ge 99\%$ (see Section \ref{sec:clustering_simulations}): for fixed $n$, increasing $d$ further does not produce benefits; on the contrary, dimensionality causes a worsening in the classification abilities. Results with $d=3$ are comparable with the ones of the k--NN and the nonparametric approach. As expected, due to the non--spherical nature of data, the GLM approach produces the worst results.

\subsubsection{Analysis of real datasets} 

In what follows we analyze the performances of our SmBP{} classifier on three real well-known datasets belonging to three very different research domains: electrocardiography, growth curves and quality control. The same datasets have been used previously in \cite{jac:pre14} in an unsupervised classification framework.

The first dataset comes from the UCR Time Series Classification and
Clustering website (\url{http://www.cs.ucr.edu/~eamonn/time\_series\_data/}). 
It consists of 200 electrocardiography (ECG) curves observed at 96
discretization points and related to 2 groups of patients (see \citealp{ols01}
for more details).

The second dataset is the well-known Berkeley growth dataset (see \citealp{tud:sny54}). It contains stature measurements for 54 girls and 39 boys,
aged from 1 to 18 years, and observed in 31 (not equispaced) discretization
points. To obtain the growth curves, the original raw data are preprocessed
by fitting each individual set of discretized data with a monotone smoothing
method (see \citealp{ram:sil05}). The aim is to discriminate the curves on the
basis of gender.

The third dataset, described in detail in \cite{lev:abr:cor:mat:mol04},
comes from Danone Vitapole Paris Research Center. The aim is to detect the
quality of produced cookies in relationship with the flour kneading process. Each curve in the dataset collects the measurements of dough resistance during the kneading process at $241$ equispaced instants of time in the interval $[0,480]$ seconds. Overall, $115$ flours are analyzed: $50$ of them
have produced cookies of good quality, $25$ of medium quality and $40$ of
low quality. The goal of the analysis is to classify the functional dataset
based on the quality of cookies.

\begin{figure}[p]
	\begin{center}
		\includegraphics[height=0.28\textheight,width=0.32\textwidth]{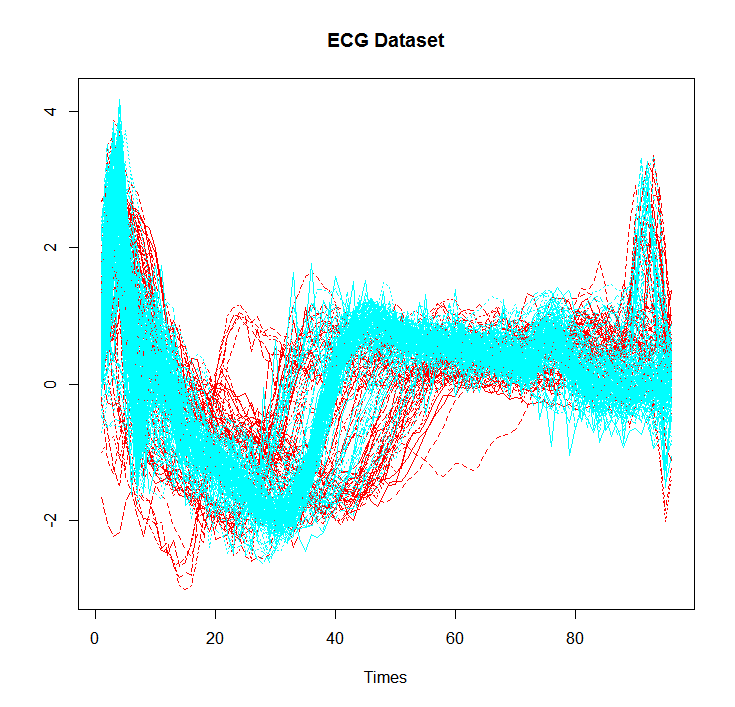} %
		\includegraphics[height=0.28\textheight,width=0.32\textwidth]{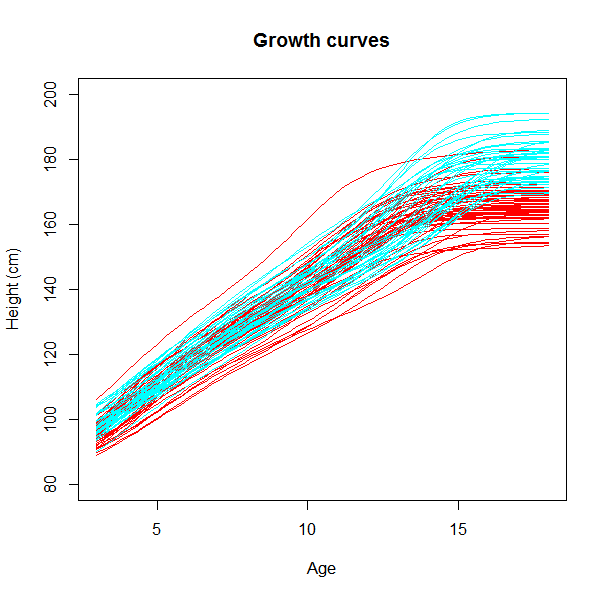} %
		\includegraphics[height=0.28\textheight,width=0.32\textwidth]{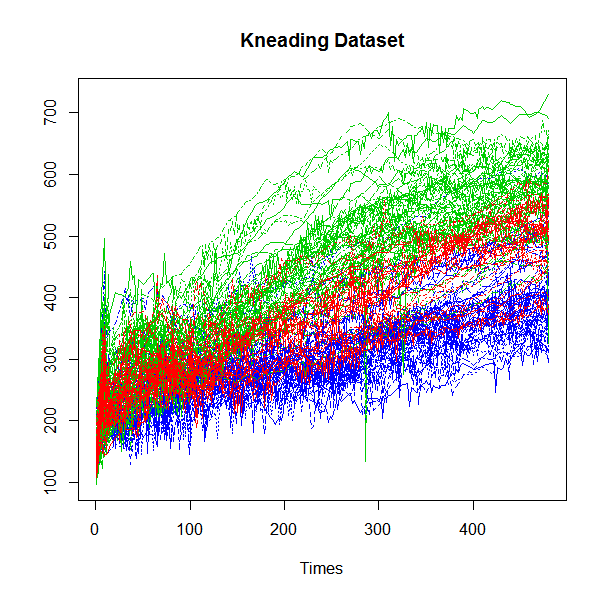}
		\\
		\includegraphics[clip=true, trim = 1cm 1cm 1cm 1cm, height=0.28\textheight,width=0.32\textwidth]{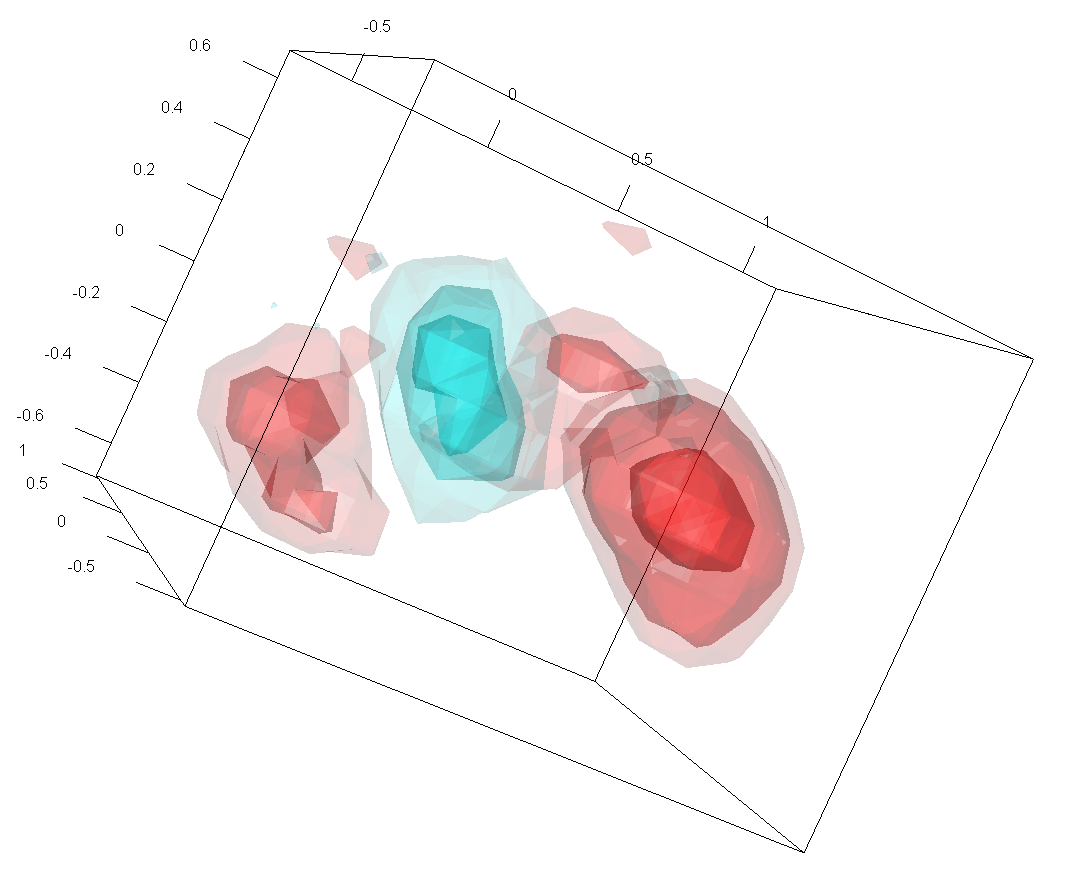}
		\includegraphics[clip=true, trim = 1cm 1cm 1cm 1cm, height=0.28\textheight,width=0.32\textwidth]{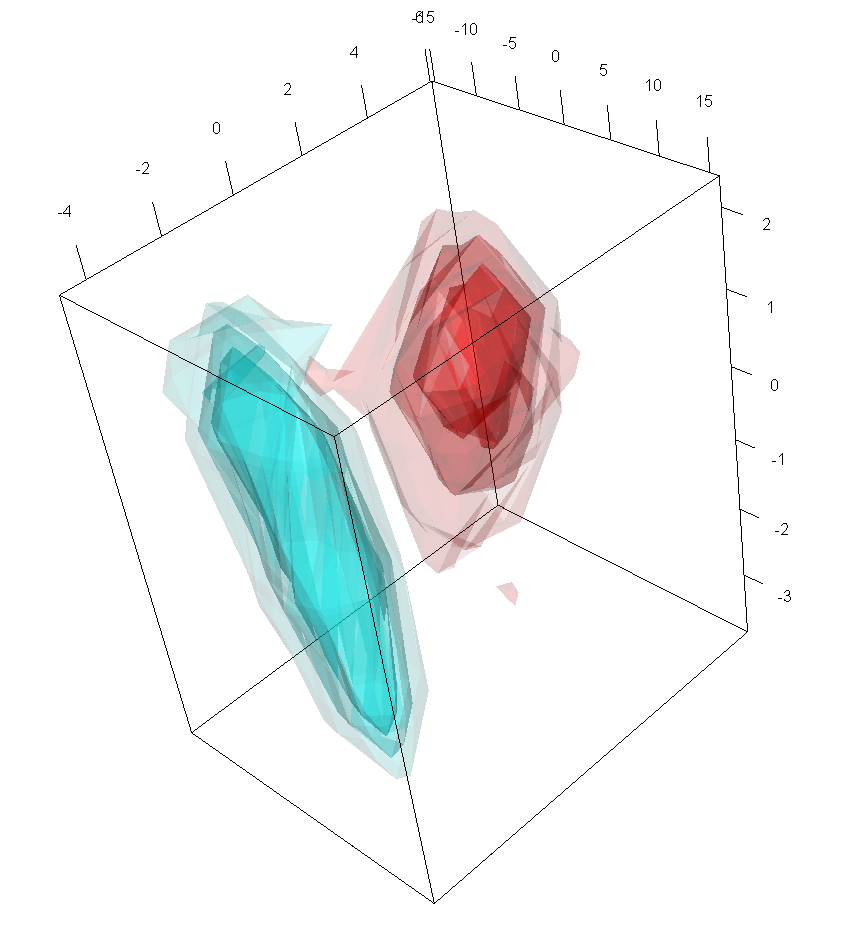} %
		\includegraphics[clip=true, trim = 1cm 1cm 1cm 1cm, height=0.28\textheight,width=0.32\textwidth]{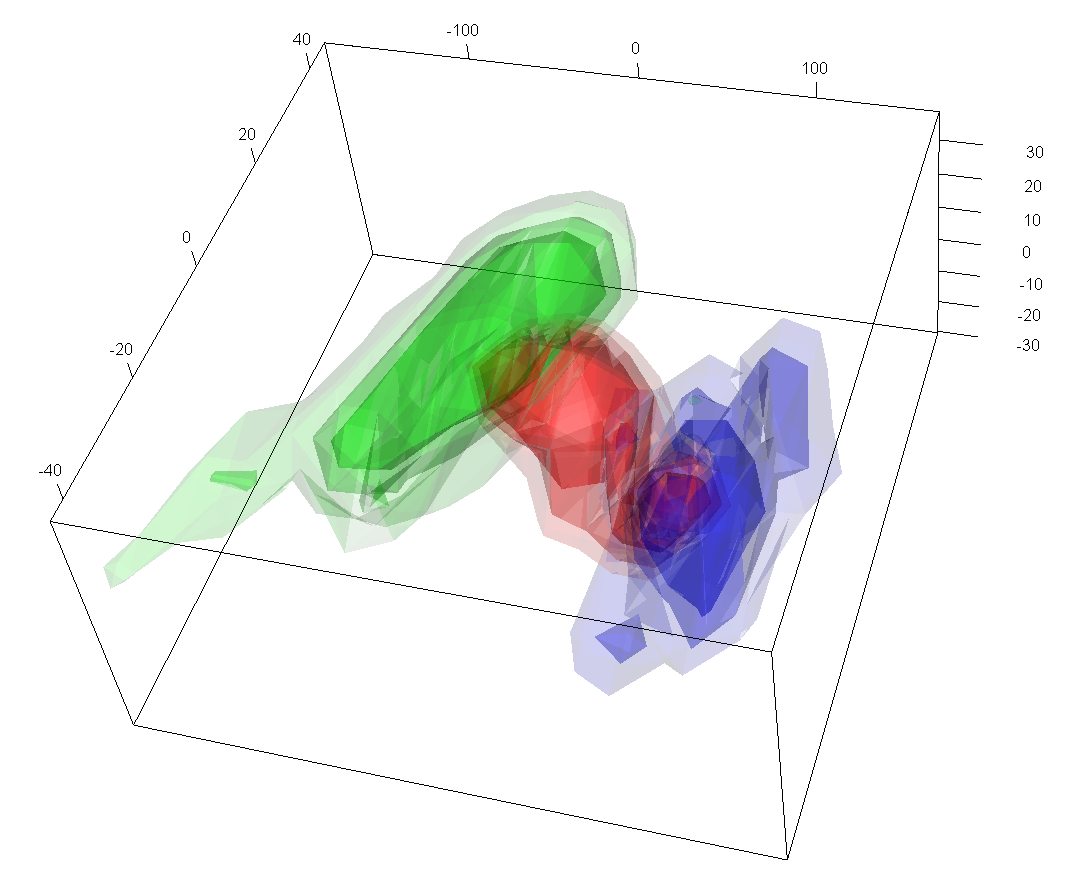}
		\\
		\includegraphics[height=0.28\textheight,width=0.32\textwidth]{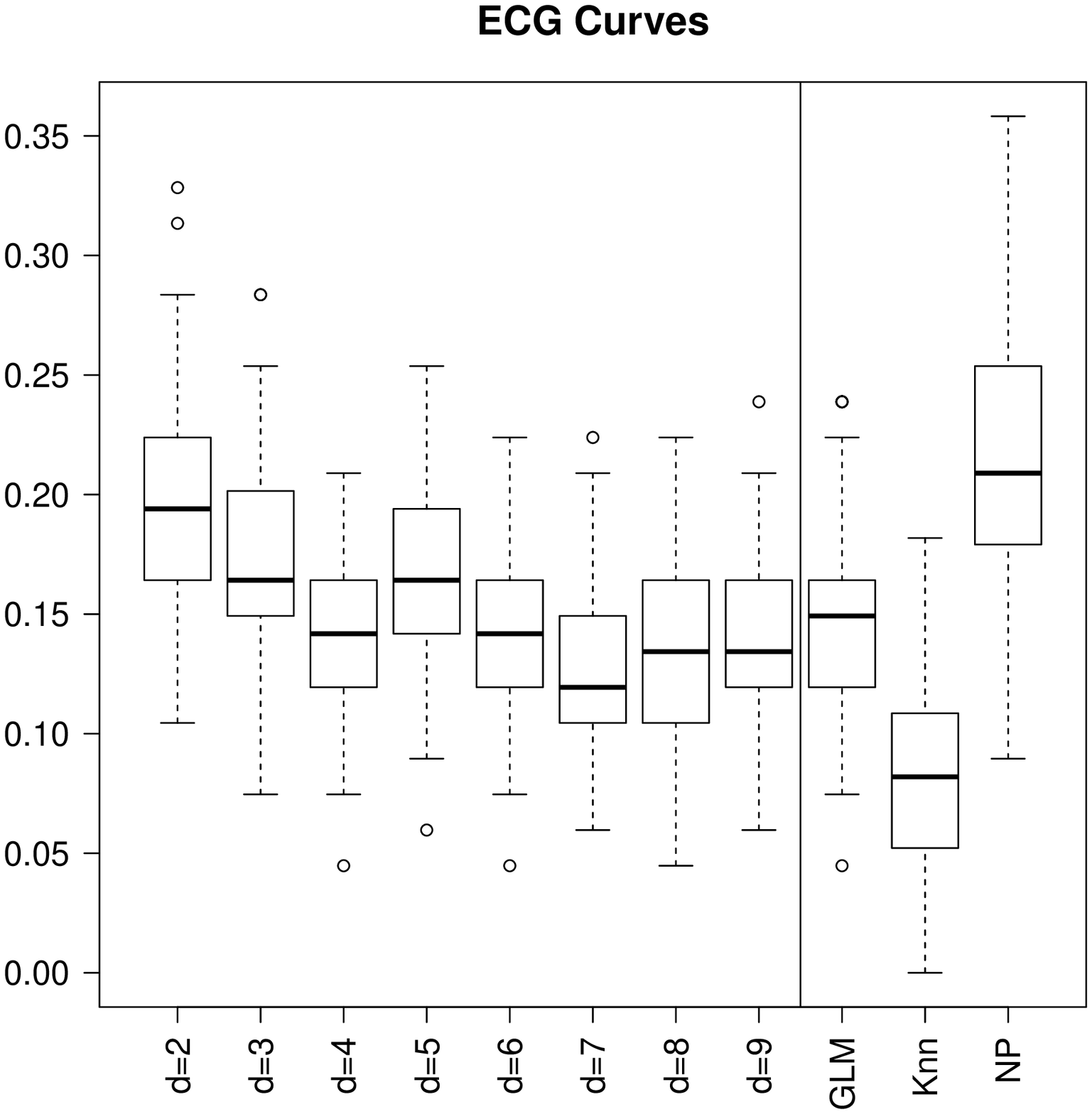} %
		\includegraphics[height=0.28\textheight,width=0.32\textwidth]{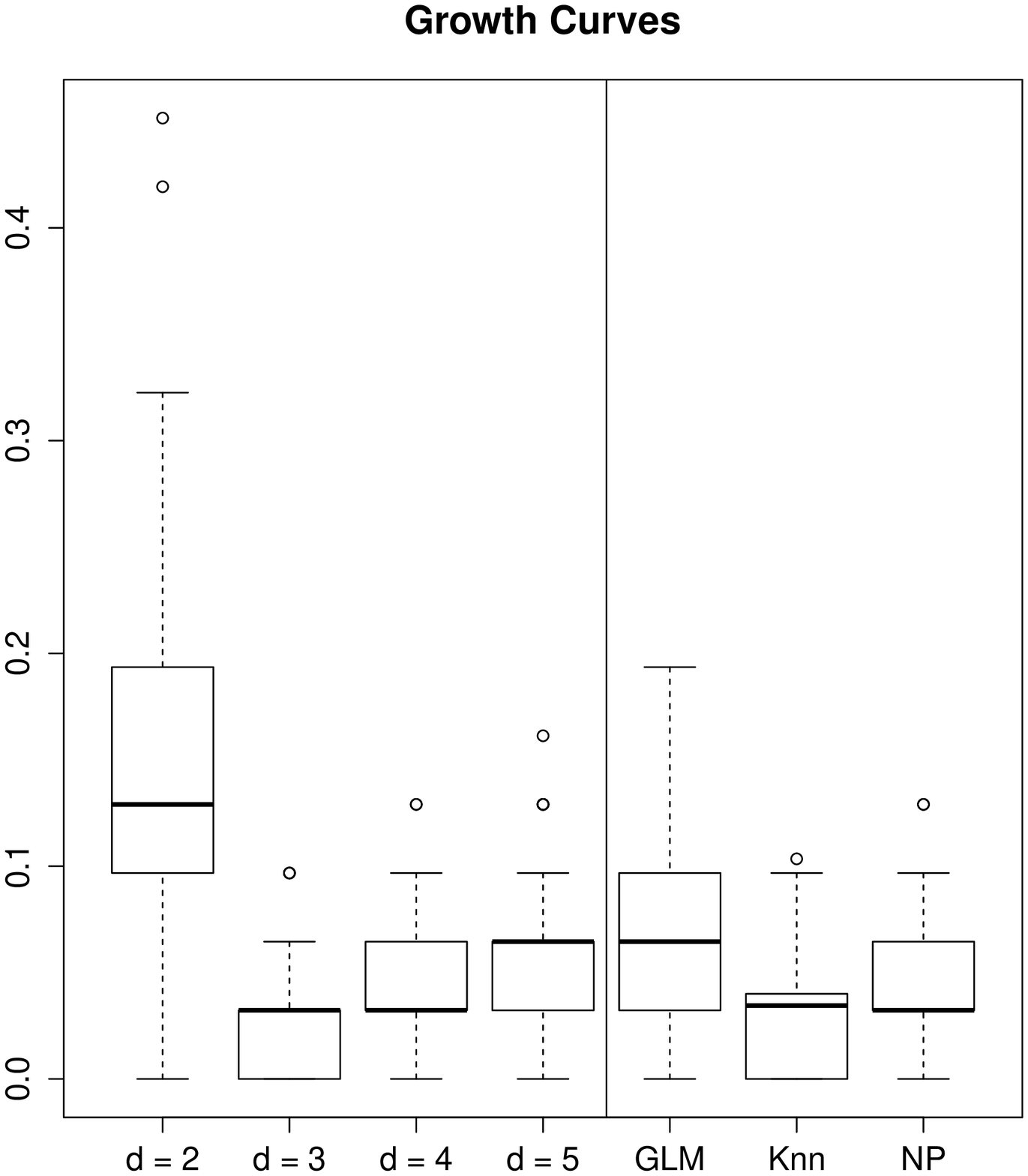} %
		\includegraphics[height=0.28\textheight,width=0.32\textwidth]{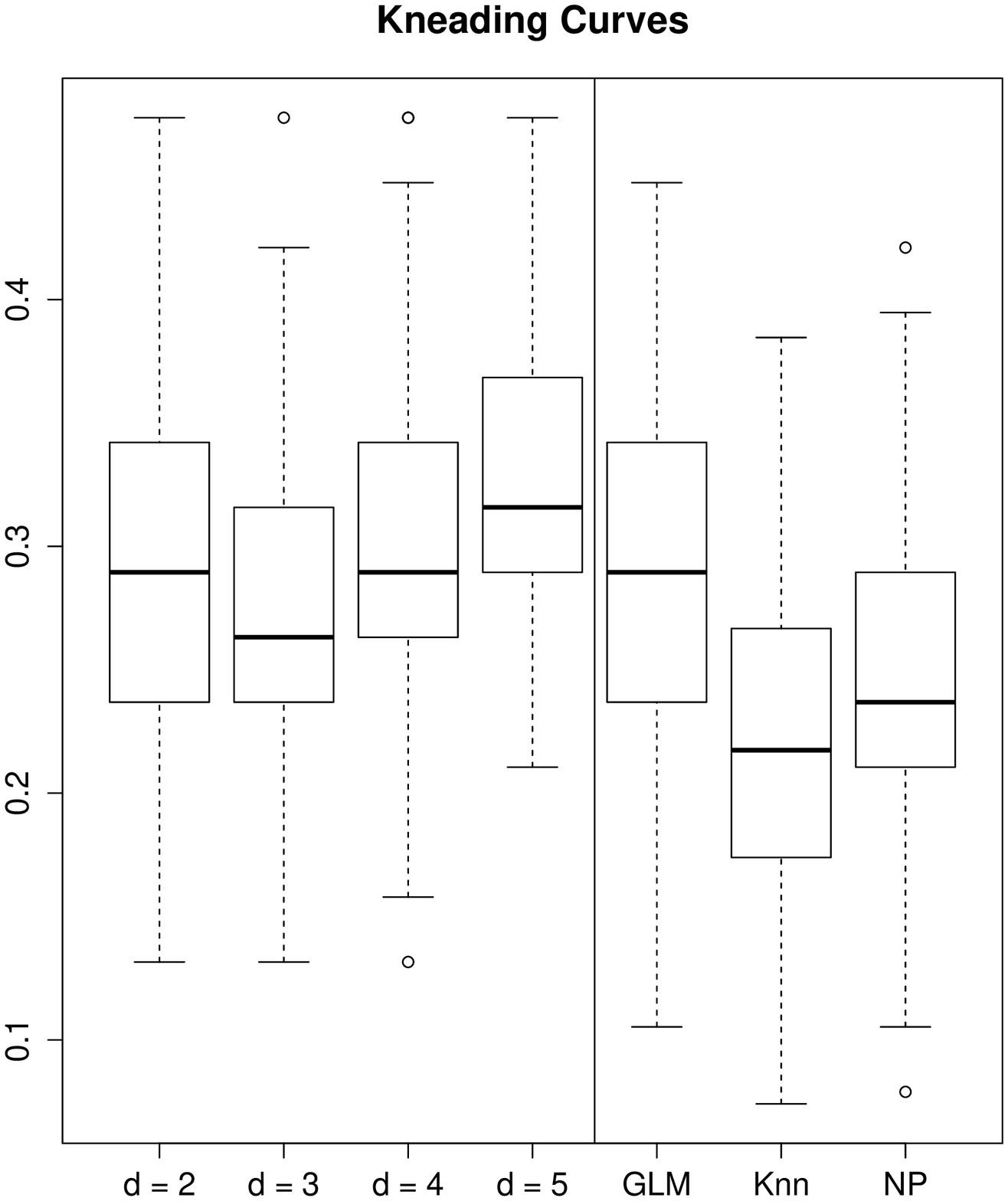}
	\end{center}
	\caption{Top to bottom: curves of the considered dataset, conditional densities of the first three PCs scores and Out-of-sample Misclassification errors over $100$ replications.
		\newline
		Case studies (left to right): ECG (2 groups), growth curves (2
		groups) and kneading process (3 groups).}
	\label{fig:RealDatasetCurves}
\end{figure}

The three datasets of curves are plotted at the top of Figure~\ref{fig:RealDatasetCurves}: in the plots, the group membership of each individual (patient, child or flour) is highlighted by using different colours. The estimated conditional pseudo--densities $\hat{f}_d(\cdot|g)$, with $d=3$, are depicted in Figure~\ref{fig:RealDatasetCurves}. Finally, Table~\ref{tab:estimated_FEV} reports the estimated FEV in the three cases.

We apply our SmBP classification method with balanced groups with the dimension $d$ which varies from $2$ to $5$. The distributions of misclassification errors for our approach and the competitors are reproduced at the bottom of Figure~\ref{fig:RealDatasetCurves}. Moreover, to allow a direct comparison, Table~\ref{tab:MisscalssErrRealData} collects the estimated mean and standard deviation of misclassification error distributions for the three real datasets: for the SmBP approach, we report the best results obtained and the corresponding dimension $d$. Unbalanced group structure (requiring the estimate of prior probabilities $\pi_g$) does not change the obtained results.
\\
The results reveal how the SmBP classifier behaves with $d$: if, in principle, misclassification errors should reduce with $d$ coherently with FEV and the mixture structure, in practice, larger values may amplify the noise due to bad estimation and, in the proposed examples, we find that a good compromise between approximation and dimensionality is reached with $d=3$ (for growth curves and kneading process dataset) and $d=4$ (for ECG dataset).

It is worth noting that our method performs rather well when compared to the other ones: despite the fact that the k--NN approach tends to produce good results uniformly in all cases, our method is always comparable, with closed results. More in detail, in the growth curves case with $d=3$, SmBP classifier is equivalent to k--NN and better than the nonparametric approach: the quality of the results is explainable by observing the estimated conditional pseudo--densities, where a thin overlapping ``grey zone'' emerges. For the ECG dataset, the best result (for $d=4$) is comparable with that obtained by nonparametric classifier. For what concerns the kneading process dataset, all of the proposed methods suffer from a relatively wide overlapping region of the estimated conditional pseudo--densities.

\begin{table}[tbp]
	\begin{center}
		\begin{tabular}{cccc} 
			\hline
			 & ECG & Growth Curves & Kneading Process \\ \hline
			SmBP & $d=4$ \ 0.148 \ (0.035) & $d=3$ \ 0.028 \ (0.024) & $d=3$ \ 0.273 \
			(0.061) \\ 
			GLM & 0.213 \ (0.037) & 0.075 \ (0.049) & 0.286 \ (0.070) \\ 
			k-NN & 0.090 \ (0.038) & 0.030 \ (0.027) & 0.221 \ (0.068) \\ 
			NP & 0.143 \ (0.036) & 0.041 \ (0.030) & 0.251 \ (0.064) \\ \hline
		\end{tabular}%
	\end{center}
	\caption{Estimated mean and standard deviation (in parentheses) of
		misclassification error distribution for the three real datasets. For the
		SmBP approach the dimension $d$ at which we obtain the best results is
		reported.}
	\label{tab:MisscalssErrRealData}
\end{table}
\begin{table}[tbp]
	\begin{center}
		\begin{tabular}{lccccc} 
			\hline
			$d$ & 1 & 2 & 3 & 4 & 5 \\
			\hline
			ECG & 38.6 & 63.4 & 73.3 & 79.2 & 83.0 \\
			Growth Curves & 81.7 & 95.9 & 98.7 & 99.5 & 99.8 \\
			Kneading Process & 89.6 & 93.2 & 95.6 & 96.3 & 96.8 \\
			\hline
		\end{tabular}
	\end{center}
	\caption{Estimated $FEV$ for the considered datasets.}\label{tab:estimated_FEV}
\end{table}


\section{Conclusions}
\label{sec:conclusion}

In this paper, an unsupervised and a supervised classification method based on the concept of SmBP mixture for Hilbert--valued process have been introduced and analyzed. The novelty lies in the use of the theoretical factorization of the SmBP due to \cite{bon:goi15} and reported in Proposition~\ref{pro:SMBP_approx}. Such a result introduces a surrogate--density for Hilbert--valued processes that, on the one hand, endorses a ``density oriented'' clustering approach for detecting the latent structure by incorporating the information on the mixture, and, on the other one, leads to define an optimal Bayes classifier in a supervised classification (discriminant) context.
From a theoretical point of view, the approaches proposed here can be seen as semi--parametric: the coefficients of the Karhunen--Lo\`{e}ve decomposition, truncated at a suitable order $d$, define the pseudo--density of a mixture model which is not entirely specified and is estimated nonparametrically. In this view, the detection of a reasonable dimension $d$ which balances the trade--off between a good approximation and the curse of dimensionality is an important task. Furthermore, dealing with joint and conditional pseudo--densities leads naturally to represent them graphically, for $d\le 3$, in order to understand the underlining mixture ``structure'' and to evaluate why classification errors may arise.
In addition to the theoretical aspects, computational issues deserve attention as well: especially for clustering, the problem of tuning parameters is deeply considered, and known tools are implemented. The use of such standard tools has shown some shortcomings strictly related to the open long--standing problem of finding a universal optimal criteria for validating clustering procedures. The issue of efficient tuning methods requires further study, which is beyond the scope of this work.


\appendix

\section{Sketch of the Proofs}
\label{sec:appendix}

This Section contains a sketch of Propositions~\ref{pro:SMBP_approx} and \ref{prop:rate_convergence}; for a detailed and theoretical discussion the interested reader can refer to \cite{bon:goi15}.

\subsection{Sketch of proof of Proposition~\ref{pro:SMBP_approx}}

At the beginning, fix $d\in\mathbb{N}$ and consider the quantities
\begin{equation*}
	S_1  = S_1(d, x)=\sum_{j\leq d} \left(\theta_j - \langle x, \xi_j\rangle \right)^{2}
	\qquad \textnormal{and} \qquad
	S = S(d,\varepsilon, x)=\frac{1}{\varepsilon^{2}} \sum_{j\geq d+1} \left(\theta_j - \langle x, \xi_j\rangle \right)^{2}.
\end{equation*}
These allow to rewrite the SmBP as follows
\begin{equation*}
	\varphi (\varepsilon ,x) = \mathbb{P} \left(\left\| X-x \right\|^2\leq \varepsilon^2 \right)= \mathbb{P}\left( S_{1}+\varepsilon ^{2}S\leq	\varepsilon^{2}\right) = \int_{0}^{1} \mathbb{P} \left( S_1 \le (1-s)\varepsilon^2 \right)dG\left( s\right),
\end{equation*}
where $G$ is the cdf of $S$. In terms of $f_d\left( \cdot \right) $, the probability density function of $\boldsymbol{\theta}=\left( \theta_{1}, \ldots ,\theta_{d}\right) ^{\prime }$, it holds
\begin{equation*}
\mathbb{P} \left( S_1 \le (1-s)\varepsilon^2 \right)=\int_{D}f_d\left( \boldsymbol{\vartheta}\right) d%
\boldsymbol{\vartheta},
\end{equation*}
with $D=\left\{ \boldsymbol{\vartheta}\in \mathbb{R}^{d}:\sum_{j\leq d}\left(\vartheta_j - x_j \right)^{2}\leq \varepsilon ^{2}\left( 1-s\right) \right\} $. 
The Taylor expansion of $f_d$ about the point $(x_1,\ldots, x_d)$ leads to the following first order approximation
\begin{equation*}
\varphi (\varepsilon ,x) \sim f_d(x) V_d \left( \varepsilon \right) \mathbb{E}\left[ \left( 1-S\right) ^{d/2}\mathbb{I}_{\left\{ S\leq 1\right\} }\right], \quad \varepsilon \rightarrow 0.
\end{equation*}
According to the type of eigenvalue decay rate, it is possible to choose $d=d(\varepsilon)$ as a function of $\varepsilon$ so that it approaches infinity when $\varepsilon$ goes to zero and such that
\begin{equation*}
S\to 0, \qquad \textnormal{and} \qquad \mathbb{E}\left[ \left( 1-S\right) ^{d/2}\mathbb{I}_{\left\{ S\leq 1\right\} }\right] \to 1.
\end{equation*}	
Finally, errors due to both the latter approximation and the Taylor expansion can be simultaneously controlled by exploiting the kind of eigenvalues decay rate; in particular, it turns out that the faster they decay, the smaller is the total error.

\subsection{Sketch of proof of Proposition~\ref{prop:rate_convergence}}

Consider $f_{d,n}$, the pseudo--estimator for $f_d$, given by
\begin{equation*} 
f_{d,n}\left( \Pi_{d}x\right) = \frac{1}{n}\sum_{i=1}^{n}K_{H_{n}}\left( \left\Vert \Pi_{d}\left( X_{i}-x\right) \right\Vert \right) ,\qquad \Pi_{d}x\in \mathbb{R}^{d},
\end{equation*}
that involves the true but unknown projector operator $\Pi_{d}\left(\cdot \right) = \sum_{j=1}^d \xi_j \left\langle \xi_j, \cdot \right\rangle $. By the triangle inequality,
\begin{equation}
\mathbb{E}\left[ f_{d}\left( x\right) -\widehat{f}_{d,n}\left( x\right) \right]
^{2}\leq \mathbb{E}\left[ f_{d}\left( x\right) -f_{d,n}\left( x\right) \right]
^{2}+\mathbb{E}\left[ f_{d,n}\left( x\right) -\widehat{f}_{d,n}\left( x\right) %
\right] ^{2}.  \label{eq:disug_triang_media_2}
\end{equation}%
Regarding the first term on the right--hand side of \eqref{eq:disug_triang_media_2}, it is well known in the literature (see for instance \citealp{wan:jon95}) that, under Assumptions (B.\ref{ass:B1})--(B.\ref{ass:B4}) and taking the optimal bandwidth \eqref{eq:optimal bandwidth}, one gets the minimax rate: 
\begin{equation*}
\mathbb{E}\left[ f_{d}\left( x\right) -f_{d,n}\left( x\right) \right]
^{2}=O\left( n^{-2p/\left( 2p+d\right) }\right)
\end{equation*}
uniformly in $\mathbb{R}^{d}$. Regarding the second addend on the right--hand side of \eqref{eq:disug_triang_media_2}, since $H_{n}=h_{n}^{2}I$, it holds
\begin{equation*}
(nh^d_n)^2\mathbb{E}\left[ f_{d,n}\left( x\right) -\widehat{f}_{d,n}\left( x\right) \right] ^{2} = \mathbb{E}\left[ \sum_{i=1}^{n}\left( K\left( \frac{V_{i}}{h_{n}}%
\right) -K\left( \frac{\widehat{V}_{i}}{h_{n}}\right) \right) \right] ^{2}
\end{equation*}
where $ V_{i}=\left\Vert \Pi _{d}\left( X_{i}-x\right) \right\Vert$, $\widehat{V}_{i}=\left\Vert \widehat{\Pi }_{d}\left( X_{i}-x\right)\right\Vert$. Consider the events $A_{i}=\left\{ V_{i}\leq h_{n}\right\}$, $B_{i}=\left\{\widehat{V}_{i}\leq h_{n}\right\}$; we get that
\begin{align*}
(nh^d_n)^2\mathbb{E}\left[ f_{d,n}\left( x\right) -\widehat{f}_{d,n}\left( x\right) \right] ^{2} \leq 
& 2\mathbb{E}\left[ \sum_{i=1}^{n}\left( K\left( \frac{V_{i}}{h_{n}}\right) -K\left( \frac{\widehat{V}_{i}}{h_{n}}\right) \right) \mathbb{I}_{_{A_{i}\cap B_{i}}}\right] ^{2}  \notag 
\\
& +4\mathbb{E}\left[ \left( \sum_{i=1}^{n}K\left( \frac{V_{i}}{h_{n}}\right)  \mathbb{I}_{_{A_{i}\cap \overline{B}_{i}}}\right) ^{2}+\left( \sum_{i=1}^{n}K\left( \frac{\widehat{V}_{i}}{h_{n}}\right) \mathbb{I}_{_{_{ \overline{A}_{i}\cap B_{i}}}}\right) ^{2}\right] .
\end{align*}
Under Assumptions~(B.\ref{ass:B1})--(B.\ref{ass:B4}) and after some computations, it is possible to show that, for any $d\geq 1$ and as $n\rightarrow \infty $ %
\begin{equation*}
\frac{1}{\left( nh_{n}^{d}\right) ^{2}} 
\mathbb{E}\left[ \sum_{i=1}^{n}\left(K\left( \frac{V_{i}}{h_{n}}\right) -K\left( \frac{\widehat{V}_{i}}{h_{n}}\right) \right) \mathbb{I}_{_{A_{i}\cap B_{i}}}\right] ^{2}
\leq C\frac{h_{n}^{2\left( d-1\right) }}{n},  
\end{equation*}%
whereas
\begin{equation*} 
\frac{1}{\left( nh_{n}^{d}\right) ^{2}} 
\mathbb{E}\left[ \left( \sum_{i=1}^{n}K\left( \frac{V_{i}}{h_{n}}\right) \mathbb{I}_{_{A_{i}\cap \overline{B}_{i}}}\right) ^{2}+
\left(\sum_{i=1}^{n}K\left( \frac{\widehat{V}_{i}}{h_{n}}\right) \mathbb{I}_{_{_{\overline{A}_{i}\cap B_{i}}}}\right) ^{2}\right] 
\leq C\left( \frac{1}{nh_{n}^{2\left( 2d+1\right) }}\right) ^{1/3}, 
\end{equation*}
and, hence,
\begin{equation} \label{eq:media2_bound}
	\mathbb{E}\left[ f_{d,n}\left( x\right) -\widehat{f}_{d,n}\left( x\right) \right] ^{2}=O\left( \frac{h_{n}^{2\left(
		d-1\right) }}{n}\right) +O\left( \left( \frac{1}{nh_{n}^{2\left( 2d+1\right)
	}}\right) ^{1/3}\right).
\end{equation}
For any $d\geq 1$, a direct computation shows that, taking the optimal bandwith \eqref{eq:optimal bandwidth} and $p>\left( 3d+2\right) /2$, the bounds in equation~\eqref{eq:media2_bound} are definitively negligible compared to the ``optimal bound'' $n^{-2p/\left(2p+d\right) }$.

{\small\paragraph{Acknowledgements}

The authors wish to thank an Associate Editor and two anonymous referees for their advice and suggestions that have allowed to improve the presentation of the paper.
The authors even wish to thank H.~H.~Bock, who provided clear comments and references about density based clustering, E.~Keogh for providing the ECG dataset, C.~Preda for providing the Danone Vitapole dataset, and A.~Schwartz, V.~Ventura and S.~Todorova for providing the neuronal experiment dataset. We thank F.~Centrone and M.~Karathanasis for their careful proofreading of the manuscript; in any case, every mistake is ascribed entirely only to the authors.
The authors were partially funded by the Gruppo Nazionale per l'Analisi Matematica, la Probabilit\`{a} e le loro Applicazioni (GNAMPA) of the Istituto Nazionale di Alta Matematica (INdAM).}

\DeclareRobustCommand{\VAN}[3]{#2}
\bibliography{SmBProb_biblio}

\begin{thebibliography}{38}
\expandafter\ifx\csname natexlab\endcsname\relax\def\natexlab#1{#1}\fi
\providecommand{\url}[1]{\texttt{#1}}
\providecommand{\href}[2]{#2}
\providecommand{\path}[1]{#1}
\providecommand{\DOIprefix}{doi:}
\providecommand{\ArXivprefix}{arXiv:}
\providecommand{\URLprefix}{URL: }
\providecommand{\Pubmedprefix}{pmid:}
\providecommand{\doi}[1]{\href{http://dx.doi.org/#1}{\path{#1}}}
\providecommand{\Pubmed}[1]{\href{pmid:#1}{\path{#1}}}
\providecommand{\bibinfo}[2]{#2}
\ifx\xfnm\relax \def\xfnm[#1]{\unskip,\space#1}\fi
\bibitem[{Abraham et~al.(2003)Abraham, Biau and Cadre}]{abr:bia:cad03}
\bibinfo{author}{Abraham, C.}, \bibinfo{author}{Biau, G.},
  \bibinfo{author}{Cadre, B.}, \bibinfo{year}{2003}.
\newblock \bibinfo{title}{Simple estimation of the mode of a multivariate
  density}.
\newblock \bibinfo{fjournal}{The Canadian Journal of Statistics. La Revue
  Canadienne de Statistique} \bibinfo{volume}{31}, \bibinfo{pages}{23--34}.
\bibitem[{Azzalini and Torelli(2007)}]{azz:tor07}
\bibinfo{author}{Azzalini, A.}, \bibinfo{author}{Torelli, N.},
  \bibinfo{year}{2007}.
\newblock \bibinfo{title}{Clustering via nonparametric density estimation}.
\newblock \bibinfo{fjournal}{Statistics and Computing} \bibinfo{volume}{17},
  \bibinfo{pages}{71--80}.
\bibitem[{Biernacki et~al.(2006)Biernacki, Celeux, Govaert and
  Langrognet}]{bie:cel:gov:lan06}
\bibinfo{author}{Biernacki, C.}, \bibinfo{author}{Celeux, G.},
  \bibinfo{author}{Govaert, G.}, \bibinfo{author}{Langrognet, F.},
  \bibinfo{year}{2006}.
\newblock \bibinfo{title}{Model-based cluster and discriminant analysis with
  the mixmod software}.
\newblock \bibinfo{fjournal}{Computational Statistics \& Data Analysis}
  \bibinfo{volume}{51}, \bibinfo{pages}{587--600}.
\bibitem[{Bock et~al.(2013)Bock, Ingrassia and Vermunt}]{boc:ing:ver13}
\bibinfo{author}{Bock, H.H.}, \bibinfo{author}{Ingrassia, S.},
  \bibinfo{author}{Vermunt, J.K.}, \bibinfo{year}{2013}.
\newblock \bibinfo{title}{Special issue on ``{M}odel-based clustering and
  classification''}.
\newblock \bibinfo{fjournal}{Advances in Data Analysis and Classification.}
  \bibinfo{volume}{7}.
\bibitem[{Bock et~al.(2014)Bock, Ingrassia and Vermunt}]{boc:ing:ver14}
\bibinfo{author}{Bock, H.H.}, \bibinfo{author}{Ingrassia, S.},
  \bibinfo{author}{Vermunt, J.K.}, \bibinfo{year}{2014}.
\newblock \bibinfo{title}{Special issue on ``{M}odel-based clustering and
  classification'' (part 2)}.
\newblock \bibinfo{fjournal}{Advances in Data Analysis and Classification.}
  \bibinfo{volume}{8}.
\bibitem[{Bongiorno and Goia(2015)}]{bon:goi15}
\bibinfo{author}{Bongiorno, E.G.}, \bibinfo{author}{Goia, A.},
  \bibinfo{year}{2015}.
\newblock \bibinfo{title}{Some insights about the small ball probability
  factorization for {H}ilbert random elements}.
\newblock \bibinfo{fjournal}{Preprint} .
\bibitem[{Bongiorno et~al.(2014)Bongiorno, Goia, Salinelli and
  Vieu}]{bon:goi:sal:vie14}
\bibinfo{editor}{Bongiorno, E.G.}, \bibinfo{editor}{Goia, A.},
  \bibinfo{editor}{Salinelli, E.}, \bibinfo{editor}{Vieu, P.} (Eds.),
  \bibinfo{year}{2014}.
\newblock \bibinfo{title}{Contributions in infinite-dimensional statistics and
  related topics}, \bibinfo{publisher}{Societ\`{a} Editrice Esculapio}.
\bibitem[{Chen et~al.(2015)Chen, Genovese and Wasserman}]{che:gen:was15}
\bibinfo{author}{Chen, Y.C.}, \bibinfo{author}{Genovese, C.R.},
  \bibinfo{author}{Wasserman, L.}, \bibinfo{year}{2015}.
\newblock \bibinfo{title}{Asymptotic theory for density ridges}.
\newblock \bibinfo{fjournal}{The Annals of Statistics} \bibinfo{volume}{43},
  \bibinfo{pages}{1896--1928}.
\bibitem[{Dabo-Niang et~al.(2007)Dabo-Niang, Ferraty and Vieu}]{dab:fer:vie07}
\bibinfo{author}{Dabo-Niang, S.}, \bibinfo{author}{Ferraty, F.},
  \bibinfo{author}{Vieu, P.}, \bibinfo{year}{2007}.
\newblock \bibinfo{title}{On the using of modal curves for radar waveforms
  classification}.
\newblock \bibinfo{fjournal}{Computational Statistics \& Data Analysis}
  \bibinfo{volume}{51}, \bibinfo{pages}{4878--4890}.
\bibitem[{Delaigle and Hall(2010)}]{del:hal10}
\bibinfo{author}{Delaigle, A.}, \bibinfo{author}{Hall, P.},
  \bibinfo{year}{2010}.
\newblock \bibinfo{title}{Defining probability density for a distribution of
  random functions}.
\newblock \bibinfo{fjournal}{The Annals of Statistics} \bibinfo{volume}{38},
  \bibinfo{pages}{1171--1193}.
\bibitem[{Delaigle et~al.(2012)Delaigle, Hall and Bathia}]{del:hal:bat12}
\bibinfo{author}{Delaigle, A.}, \bibinfo{author}{Hall, P.},
  \bibinfo{author}{Bathia, N.}, \bibinfo{year}{2012}.
\newblock \bibinfo{title}{Componentwise classification and clustering of
  functional data}.
\newblock \bibinfo{fjournal}{Biometrika} \bibinfo{volume}{99},
  \bibinfo{pages}{299--313}.
\bibitem[{Devroye(1981)}]{dev81}
\bibinfo{author}{Devroye, L.}, \bibinfo{year}{1981}.
\newblock \bibinfo{title}{On the almost everywhere convergence of nonparametric
  regression function estimates}.
\newblock \bibinfo{fjournal}{The Annals of Statistics} \bibinfo{volume}{9},
  \bibinfo{pages}{1310--1319}.
\bibitem[{Dubes(1993)}]{dub93}
\bibinfo{author}{Dubes, R.C.}, \bibinfo{year}{1993}.
\newblock \bibinfo{title}{Cluster analysis and related issues}, in:
  \bibinfo{editor}{Chen, C.H.}, \bibinfo{editor}{Pau, L.F.},
  \bibinfo{editor}{Wang, P.S.P.} (Eds.), \bibinfo{booktitle}{Handbook of
  Pattern Recognition and Computer Vision}. \bibinfo{publisher}{World
  Scientific Publishing Co., Inc.}, \bibinfo{address}{River Edge, NJ, USA}, pp.
  \bibinfo{pages}{3--32}.
\bibitem[{Duong and Hazelton(2005)}]{duo:haz05}
\bibinfo{author}{Duong, T.}, \bibinfo{author}{Hazelton, M.L.},
  \bibinfo{year}{2005}.
\newblock \bibinfo{title}{Cross-validation bandwidth matrices for multivariate
  kernel density estimation}.
\newblock \bibinfo{fjournal}{Scandinavian Journal of Statistics. Theory and
  Applications} \bibinfo{volume}{32}, \bibinfo{pages}{485--506}.
\bibitem[{Febrero-Bande and {Oviedo de la Fuente}(2012)}]{feb:ovi12}
\bibinfo{author}{Febrero-Bande, M.}, \bibinfo{author}{{Oviedo de la Fuente},
  M.}, \bibinfo{year}{2012}.
\newblock \bibinfo{title}{Statistical computing in functional data analysis:
  The {R} package {fda.usc}}.
\newblock \bibinfo{fjournal}{Journal of Statistical Software}
  \bibinfo{volume}{51}, \bibinfo{pages}{1--28}.
\bibitem[{Ferraty et~al.(2012)Ferraty, Kudraszow and Vieu}]{fer:kud:vie12}
\bibinfo{author}{Ferraty, F.}, \bibinfo{author}{Kudraszow, N.},
  \bibinfo{author}{Vieu, P.}, \bibinfo{year}{2012}.
\newblock \bibinfo{title}{Nonparametric estimation of a surrogate density
  function in infinite-dimensional spaces}.
\newblock \bibinfo{fjournal}{Journal of Nonparametric Statistics}
  \bibinfo{volume}{24}, \bibinfo{pages}{447--464}.
\bibitem[{Ferraty and Vieu(2003)}]{fer:vie03}
\bibinfo{author}{Ferraty, F.}, \bibinfo{author}{Vieu, P.},
  \bibinfo{year}{2003}.
\newblock \bibinfo{title}{Curves discrimination: a nonparametric functional
  approach}.
\newblock \bibinfo{fjournal}{Computational Statistics \& Data Analysis}
  \bibinfo{volume}{44}, \bibinfo{pages}{161--173}.
\bibitem[{Ferraty and Vieu(2006)}]{fer:vie06}
\bibinfo{author}{Ferraty, F.}, \bibinfo{author}{Vieu, P.},
  \bibinfo{year}{2006}.
\newblock \bibinfo{title}{Nonparametric functional data analysis}.
\newblock Springer Series in Statistics, \bibinfo{publisher}{Springer, New
  York}.
\bibitem[{Gasser et~al.(1998)Gasser, Hall and Presnell}]{gas:hal:pre98}
\bibinfo{author}{Gasser, T.}, \bibinfo{author}{Hall, P.},
  \bibinfo{author}{Presnell, B.}, \bibinfo{year}{1998}.
\newblock \bibinfo{title}{Nonparametric estimation of the mode of a
  distribution of random curves}.
\newblock \bibinfo{fjournal}{Journal of the Royal Statistical Society. Series
  B. Statistical Methodology} \bibinfo{volume}{60}, \bibinfo{pages}{681--691}.
\bibitem[{Gimelʹfarb et~al.(2012)Gimelʹfarb, Hancock, Imiya, Kuijper, Kudo,
  Omachi, Windeatt and Yamada}]{gim:han:imi12}
\bibinfo{author}{Gimelʹfarb, G.}, \bibinfo{author}{Hancock, E.},
  \bibinfo{author}{Imiya, A.}, \bibinfo{author}{Kuijper, A.},
  \bibinfo{author}{Kudo, M.}, \bibinfo{author}{Omachi, S.},
  \bibinfo{author}{Windeatt, T.}, \bibinfo{author}{Yamada, K.},
  \bibinfo{year}{2012}.
\newblock \bibinfo{title}{Structural, Syntactic, and Statistical Pattern
  Recognition: Joint IAPR International Workshop, SSPR\&SPR 2012, Hiroshima,
  Japan, November 7-9, 2012. Proceedings}.
\newblock \bibinfo{publisher}{Springer}.
\bibitem[{Goia(2012)}]{goi12}
\bibinfo{author}{Goia, A.}, \bibinfo{year}{2012}.
\newblock \bibinfo{title}{A functional linear model for time series prediction
  with exogenous variables}.
\newblock \bibinfo{fjournal}{Statistics \& Probability Letters}
  \bibinfo{volume}{82}, \bibinfo{pages}{1005--1011}.
\bibitem[{Goia et~al.(2010)Goia, May and Fusai}]{goi:may:fus10}
\bibinfo{author}{Goia, A.}, \bibinfo{author}{May, C.}, \bibinfo{author}{Fusai,
  G.}, \bibinfo{year}{2010}.
\newblock \bibinfo{title}{Functional clustering and linear regression for peak
  load forecasting}.
\newblock \bibinfo{fjournal}{International Journal of Forecasting}
  \bibinfo{volume}{26}, \bibinfo{pages}{700--711}.
\bibitem[{Horv{\'a}th and Kokoszka(2012)}]{hor:kok12}
\bibinfo{author}{Horv{\'a}th, L.}, \bibinfo{author}{Kokoszka, P.},
  \bibinfo{year}{2012}.
\newblock \bibinfo{title}{Inference for functional data with applications}.
  volume \bibinfo{volume}{200}.
\newblock \bibinfo{publisher}{Springer Science \& Business Media}.
\bibitem[{Jacques and Preda(2014)}]{jac:pre14}
\bibinfo{author}{Jacques, J.}, \bibinfo{author}{Preda, C.},
  \bibinfo{year}{2014}.
\newblock \bibinfo{title}{Model-based clustering for multivariate functional
  data}.
\newblock \bibinfo{fjournal}{Computational Statistics \& Data Analysis}
  \bibinfo{volume}{71}, \bibinfo{pages}{92--106}.
\bibitem[{James and Hastie(2001)}]{jam:has01}
\bibinfo{author}{James, G.M.}, \bibinfo{author}{Hastie, T.J.},
  \bibinfo{year}{2001}.
\newblock \bibinfo{title}{Functional linear discriminant analysis for
  irregularly sampled curves}.
\newblock \bibinfo{fjournal}{Journal of the Royal Statistical Society: Series B
  (Statistical Methodology)} \bibinfo{volume}{63}, \bibinfo{pages}{533--550}.
\bibitem[{James and Sugar(2003)}]{jam:sug03}
\bibinfo{author}{James, G.M.}, \bibinfo{author}{Sugar, C.A.},
  \bibinfo{year}{2003}.
\newblock \bibinfo{title}{Clustering for sparsely sampled functional data}.
\newblock \bibinfo{fjournal}{Journal of the American Statistical Association}
  \bibinfo{volume}{98}, \bibinfo{pages}{397--408}.
\bibitem[{L{\'e}v{\'e}der et~al.(2004)L{\'e}v{\'e}der, Abraham, Cornillon,
  Matzner-Lober and Molinari}]{lev:abr:cor:mat:mol04}
\bibinfo{author}{L{\'e}v{\'e}der, C.}, \bibinfo{author}{Abraham, C.},
  \bibinfo{author}{Cornillon, P.}, \bibinfo{author}{Matzner-Lober, E.},
  \bibinfo{author}{Molinari, N.}, \bibinfo{year}{2004}.
\newblock \bibinfo{title}{Discrimination de courbes de p{\'e}trissage}.
\newblock \bibinfo{fjournal}{Chimiom{\'e}trie} , \bibinfo{pages}{37--43}.
\bibitem[{Liu et~al.(2010)Liu, Chen, Maisog and Luta}]{liu:che:mai:lut10}
\bibinfo{author}{Liu, J.}, \bibinfo{author}{Chen, Y.}, \bibinfo{author}{Maisog,
  J.M.}, \bibinfo{author}{Luta, G.}, \bibinfo{year}{2010}.
\newblock \bibinfo{title}{A new point containment test algorithm based on
  preprocessing and determining triangles}.
\newblock \bibinfo{fjournal}{Computer-Aided Design} \bibinfo{volume}{42},
  \bibinfo{pages}{1143 -- 1150}.
\bibitem[{Olszewski(2001)}]{ols01}
\bibinfo{author}{Olszewski, R.T.}, \bibinfo{year}{2001}.
\newblock \bibinfo{title}{Generalized feature extraction for structural pattern
  recognition in time-series data. {P}h{D} {T}hesis}.
\bibitem[{Ramsay and Silverman(2005)}]{ram:sil05}
\bibinfo{author}{Ramsay, J.O.}, \bibinfo{author}{Silverman, B.W.},
  \bibinfo{year}{2005}.
\newblock \bibinfo{title}{Functional data analysis}.
\newblock Springer Series in Statistics. \bibinfo{edition}{second} ed.,
  \bibinfo{publisher}{Springer, New York}.
\bibitem[{Rinaldo et~al.(2012)Rinaldo, Singh, Nugent and
  Wasserman}]{rin:sin:nug:was12}
\bibinfo{author}{Rinaldo, A.}, \bibinfo{author}{Singh, A.},
  \bibinfo{author}{Nugent, R.}, \bibinfo{author}{Wasserman, L.},
  \bibinfo{year}{2012}.
\newblock \bibinfo{title}{Stability of density-based clustering}.
\newblock \bibinfo{fjournal}{Journal of Machine Learning Research}
  \bibinfo{volume}{13}, \bibinfo{pages}{905--948}.
\bibitem[{Sager(1979)}]{sag79}
\bibinfo{author}{Sager, T.W.}, \bibinfo{year}{1979}.
\newblock \bibinfo{title}{An iterative method for estimating a multivariate
  mode and isopleth}.
\newblock \bibinfo{fjournal}{Journal of the American Statistical Association}
  \bibinfo{volume}{74}, \bibinfo{pages}{329--339}.
\bibitem[{Shin(2008)}]{shi08}
\bibinfo{author}{Shin, H.}, \bibinfo{year}{2008}.
\newblock \bibinfo{title}{An extension of fisher's discriminant analysis for
  stochastic processes}.
\newblock \bibinfo{fjournal}{Journal of Multivariate Analysis}
  \bibinfo{volume}{99}, \bibinfo{pages}{1191--1216}.
\bibitem[{Silverman(1986)}]{sil86}
\bibinfo{author}{Silverman, B.W.}, \bibinfo{year}{1986}.
\newblock \bibinfo{title}{Density estimation for statistics and data analysis}.
\newblock Monographs on Statistics and Applied Probability,
  \bibinfo{publisher}{Chapman \& Hall, London}.
\bibitem[{Todorova et~al.(2014)Todorova, Sadtler, Batista, Chase and
  Ventura}]{tod:sad:bat:cha:ven14}
\bibinfo{author}{Todorova, S.}, \bibinfo{author}{Sadtler, P.},
  \bibinfo{author}{Batista, A.}, \bibinfo{author}{Chase, S.},
  \bibinfo{author}{Ventura, V.}, \bibinfo{year}{2014}.
\newblock \bibinfo{title}{To sort or not to sort: the impact of spike-sorting
  on neural decoding performance}.
\newblock \bibinfo{fjournal}{Journal of neural engineering}
  \bibinfo{volume}{11}, \bibinfo{pages}{056005}.
\bibitem[{Tuddenham and Snyder(1954)}]{tud:sny54}
\bibinfo{author}{Tuddenham, R.}, \bibinfo{author}{Snyder, M.},
  \bibinfo{year}{1954}.
\newblock \bibinfo{title}{Physical growth of california boys and girls from
  birth to age 18}.
\newblock \bibinfo{fjournal}{California Publications on Child Development}
  \bibinfo{volume}{1}, \bibinfo{pages}{183--364}.
\bibitem[{Wand and Jones(1995)}]{wan:jon95}
\bibinfo{author}{Wand, M.P.}, \bibinfo{author}{Jones, M.C.},
  \bibinfo{year}{1995}.
\newblock \bibinfo{title}{Kernel smoothing}. volume~\bibinfo{volume}{60} of
  \textit{\bibinfo{series}{Monographs on Statistics and Applied Probability}}.
\newblock \bibinfo{publisher}{Chapman and Hall, Ltd., London}.
\bibitem[{Xu and Wunsch~II(2005)}]{xu:wun05}
\bibinfo{author}{Xu, R.}, \bibinfo{author}{Wunsch~II, D.},
  \bibinfo{year}{2005}.
\newblock \bibinfo{title}{Survey of clustering algorithms}.
\newblock \bibinfo{fjournal}{Neural Networks, IEEE Transactions on}
  \bibinfo{volume}{16}, \bibinfo{pages}{645--678}.

\end{thebibliography}

\end{document}